\documentclass[manuscript]{aastex}
\usepackage{epstopdf,natbib}
\usepackage{fullpage}
\usepackage{epstopdf,natbib}
\usepackage{graphicx} 
\usepackage{wrapfig}
\usepackage{amsmath}

\begin{document}

\title{NEOWISE Reactivation Mission Year Two: Asteroid Diameters and Albedos}
\author{C. R. Nugent\altaffilmark{1}, 
A. Mainzer\altaffilmark{2}, 
J. Bauer\altaffilmark{2}, 
R. M. Cutri\altaffilmark{1},
E. A. Kramer\altaffilmark{2}, 
T. Grav\altaffilmark{3}, 
J. Masiero\altaffilmark{2}, 
S. Sonnett\altaffilmark{2}, and 
E. L. Wright\altaffilmark{4}}
\altaffiltext{1}{Infrared Processing and Analysis Center, California Institute of Technology, Pasadena, CA 
91125, USA}
\altaffiltext{2}{Jet Propulsion Laboratory, California Institute of Technology, Pasadena, CA 91109 USA}
\altaffiltext{3}{Planetary Science Institute, Tucson, AZ}
\altaffiltext{4}{Department of Physics and Astronomy, University of California, Los Angeles, CA 90095, USA}

\begin{abstract} 

The Near-Earth Object Wide-Field Infrared Survey Explorer (NEOWISE) mission continues to detect, track, and characterize minor planets. We present diameters and albedos calculated from observations taken during the second year since the spacecraft was reactivated in late 2013. These include 207 near-Earth asteroids and 8,885 other asteroids. $84\%$ of the near-Earth asteroids did not have previously measured diameters and albedos by the NEOWISE mission. Comparison of sizes and albedos calculated from NEOWISE measurements with those measured by occultations, spacecraft, and radar-derived shapes shows accuracy consistent with previous NEOWISE publications. Diameters and albedos fall within $ \pm \sim20\%$ and $\pm\sim40\%$, 1-sigma, respectively, of those measured by these alternate techniques. NEOWISE continues to preferentially discover near-Earth objects which are large ($>100$ m), and have low albedos. 

\end{abstract}

\section{Introduction}

Observing asteroids at infrared wavelengths is an effective method for calculating diameters for large numbers of asteroids. Since asteroid albedos can vary by approximately an order of magnitude, sizes estimated from reflected visible light fluxes alone have large uncertainties. Combining diameters calculated from infrared fluxes with visible magnitudes yields albedo measurements. Together, diameter and albedo measurements are basic physical characterizations that enable further investigations, including studies of asteroid families \citep{Masiero.2013a, Walsh.2013a, Carruba.2013a, Milani.2014a,AsteroidsIVFam, MasieroEuphrosyne} and size-frequency distributions \citep{1979aste.book..783Z,1982Sci...216.1405G,2002AJ....123.1056T,2002Icar..158..146B,Mainzer2011d,GravHilda,GravTrojans, GerbsCentaur}.

We present diameters and albedos of asteroids from the second  year of the NEOWISE mission following the reactivation of the spacecraft from hibernation in late 2013. Diameters and albedos of asteroids from the first year of the NEOWISE mission following reactivation are given in \citet{Nugent15}. NEOWISE is a space-based infrared telescope that obtains an image of the sky every eleven seconds simultaneously in two bands, W1 (3.4 $\mu$m) and W2 (4.6 $\mu$m). From its sun-synchronous orbit around Earth, NEOWISE observes the entire static sky every six months. The original mission, WISE, is described in detail in \citet{Wright10WISE}, and the NEOWISE enhancement to the mission is described in \citet{Mainzer2011a}. After successfully completing its prime mission in 2011, the WISE spacecraft was placed into hibernation for 32 months before being reactivated and renamed NEOWISE in late 2013. The NEOWISE reactivation mission is described in \citet{14MainzerRestart}.

The goals of the NEOWISE mission are to discover, track, and characterize minor planets. Images and extracted source lists from all phases of the WISE and NEOWISE missions have been delivered to the public via the Infrared Science Archive \citep{Cutri12,2015Cutri}, NASA's designated archive for infrared astronomical data. 

During the initial portion of the mission, NEOWISE employed four channels; 3.4, 4.6, 12, and 22 $\mu$m. The longest two wavelength channels required cooling to $<8$ K using a dual-stage solid hydrogen cryostat. Diameters and albedos for a variety of small body populations were calculated using this fully cryogenic portion of the mission (see Table \ref{tab:cryo}).  As the cryogen was depleted, the 12 and 22 $\mu$m channels became inoperative; after this, the mission continued for several months using only its 3.4 and 4.6 $\mu$m channels. A summary of near-Earth asteroid (NEAs) and main belt asteroid (MBA) albedos and diameters calculated during various phases of the mission is given in Table \ref{tabref}. These measurements have also been submitted to NASA's Planetary Data System. Thermal model calibration results, including comparison of cryogenic WISE/NEOWISE-derived diameters to other observations, are given in \citet{Mainzer2011b, Mainzer2011c}.

\begin{deluxetable}{rl}
\tabletypesize{\scriptsize}
\tablecaption{Diameters and albedos for various small body populations, calculated from fully cryogenic (3.4, 4.6, 12 and 22 $\mu$m bands) NEOWISE mission data.}
\tablewidth{0pt}
\tablehead{
\colhead{Population} & \colhead{Associated reference} } 
\startdata
Near-Earth asteroids     & \citet{Mainzer2011d}, \citet{2014MainzerTiny} \\
Main belt asteroids      & \citet{Masiero11}    \\
Active main belt objects & \citet{Bauer12}      \\
Trojans                  & \citet{GravTrojans2} \\
Hildas                   & \citet{GravHilda}    \\
Irregular satellites     & \citet{GravIrregular}\\
Centaurs                 & \citet{GerbsCentaur} \\
Comets                   & \citet{Bauer11,Bauer12b,Bauer15},                                                                         
\enddata
\label{tab:cryo}
\end{deluxetable}

\begin{deluxetable}{rlll}
\tabletypesize{\scriptsize}
\tablecaption{Previously published papers containing NEA and MBA diameters and albedos calculated from NEOWISE mission data.}
\tablewidth{0pt}
\tablehead{
\colhead{Mission Phase} & \colhead{Detection bands ($\mu$m)} & \colhead{Reference for NEAs} & \colhead{Reference for MBAs} } 
\startdata
Fully cryogenic           & 3.4, 4.6, 12, 22    & \citet{Mainzer2011d},  & \citet{Masiero11},  \\
                          &                     & \citet{2014MainzerTiny} & \citet{Masiero14} \\
3-Band and Post cryogenic & 3.4, 4.6, (some 12) & \citet{Mainzer12}    & \citet{Masiero12} \\
Year 1 of reactivation    & 3.4, 4.6            & \citet{Nugent15}     & \citet{Nugent15} \\                                                                                    
\enddata
\label{tabref}
\end{deluxetable}

This second year of data also provides multi-epoch observational data of uniform quality that can be used to better constrain the sizes, shapes, rotation state and thermophysical properties of the  9,092 asteroids in the reactivation Year 2 sample.

We present preliminary diameters and albedos calculated from NEOWISE Year 2 Reactivation mission observations, which spanned 13 December 2014 to 13 December 2015. Diameters and albedos calculated from NEOWISE Year 2 Reactivation mission observations will be submitted to the Planetary Data System.

\section{Discoveries and follow-up}

NEOWISE discovered 198 near-Earth asteroids and comets during Years 1 and 2 of the Reactivation mission. In addition to observing 175 NEAs that had not had diameters measured previously from NEOWISE data, the Year 2 Reactivation mission obtained thermal infrared observations at additional epochs for 32 NEAs. NEOWISE typically observes asteroids $\sim10-12$ times over $\sim 1-1.5$ days, and requires a minimum of 5 detections of a discovery candidate for submission to the Minor Planet Center (MPC). 

NEOWISE observes with a fixed observing cadence, and additional follow up observations are usually necessary to confirm that new minor planet candidates have been discovered. Since NEOWISE cannot perform targeted follow up on its own, these observations must be made by ground-based observers. Given that near-Earth objects (NEOs) are of a population of special interest, NEOWISE candidate NEOs are listed on the MPC Near-Earth Object Confirmation Page (NEOCP) to facilitate follow up. NEOWISE regularly relies on many ground-based observers for follow up, including Spacewatch, observers at the Institute for Astronomy at the University of Hawaii, the Las Cumbres Observatory Global telescope Network (LCOGT), the Magdalena Ridge Observatory, the Mt. John Observatory, and a number of amateur observers across the globe to coordinate follow up of particular objects. The NEOWISE team was granted eight hours each semester of Target of Opportunity observing time on Gemini Observatory's Gemini South telescope \citep{Hook04} as well as time on the Blanco 4m/Dark Energy Camera \citep[DECam;][]{DECam15}, and was granted Co-I status on the LCOGT NEO follow up program. Access to these facilities is vital for following up of discoveries deep in the Southern Hemisphere.

\begin{figure}[h!]
  \caption{NEOWISE relies on ground-based observatories to perform targeted follow up of candidate NEO discoveries. This histogram shows numbers of follow-up observations taken by different groups (listed by their MPC observatory codes) for NEOWISE discoveries during Year 1 and Year 2 of the Reactivation mission.
  }
  \centering
  \includegraphics[width=1.0\textwidth]{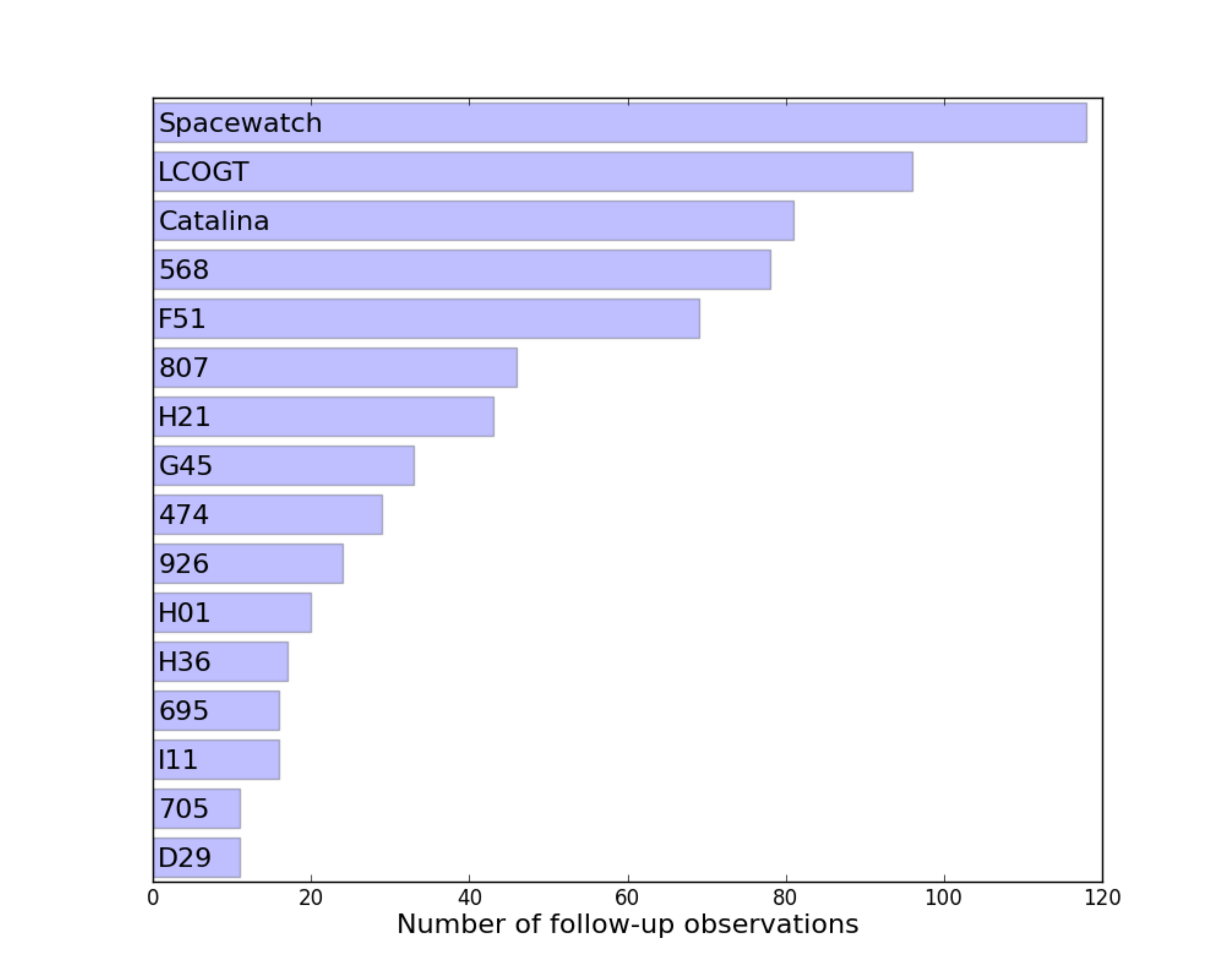}
  \label{fig:followup}
\end{figure}

Figure \ref{fig:followup} is a histogram of the observatories and campaigns that contributed the majority of follow observations occurring immediately after NEOWISE reported candidate observations, including Spacewatch \citep{2007McMillan}, with over 100 follow-up observations, the Las Cumbres Observatory Global Telescope Network \citep{Brown2013}, and the Catalina Sky Survey \citep{2015DPSCatalina}. Observers on Mauna Kea using the University of Hawaii 2.2m telescope and the Canada-France-Hawaii Telescope's Megacam imager repeatedly obtained follow-up of objects under challenging observing conditions \citep[e.g.][]{2014MPECTholen}. The Mt John University Observatory (observatory code 474) obtained valuable observations of 2015 OA$_{22}$. NEOWISE discovered this object at $-70^{\circ}$ declination, and ephemerides showed it was moving further South. The Mt John University Observatory was able to track the object to $-78^{\circ}$ declination, confirming the discovery.

NEOWISE submitted several candidate objects to the MPC that were not placed on the NEOCP based on an initial orbit determination that indicated they were not NEOs. These did not receive targeted follow-up. Fifty-seven non-NEO NEOWISE discoveries made during the Reactivation mission do not have associated orbits. Additionally, there were eight objects (main belt asteroids, Hungarias, and Mars-crossers) detected solely by NEOWISE that were given provisional designations by the MPC; all of these objects have poorly determined orbits. Without well-determined orbits, distance at observing time cannot be computed accurately, and therefore diameters were not determined for these objects.

\subsection{Comets}
\begin{figure}[h!]
  \caption{Comet C/2013 US$_{10}$ (Catalina) as seen by NEOWISE on August 28, 2015. The 2-color image maps the 3.4 $\mu$m band to the cyan, and the 4.6 $\mu$m band to the red. The image is a quarter-degree on a side and is oriented approximately with North down and East is to the left. The red appearance suggests the comet may have significant CO or CO$_2$ emission.
  }
  \centering
  \includegraphics[width=0.8\textwidth]{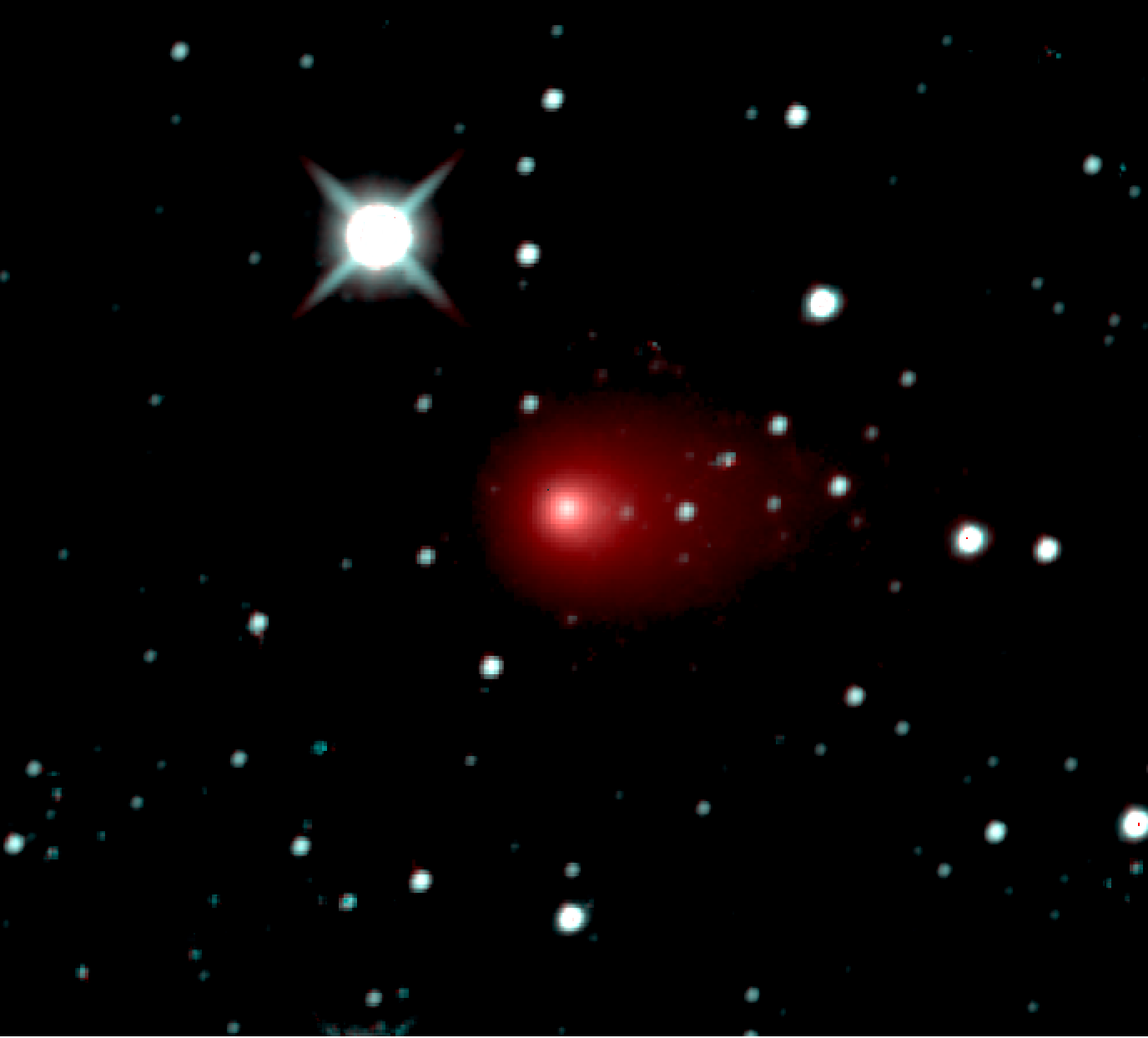}
  \label{fig:catalina}
\end{figure}

The NEOWISE Reactivation mission has detected over 100 comets, including eight discoveries (four of which were made after the end of the second year of observations). The NEOWISE spacecraft is sensitive to the presence of coma dust, as well as the CO-line (4.67 $\mu$m)  and CO$_2$-line (4.23 $\mu$m) emission from comet comae from gas species which are obscured or completely blocked by Earth's atmosphere (Figure \ref{fig:catalina}).  Analysis of the excess emission at 3.4 $\mu$m by \citet{Bauer15} provided CO+CO$_2$ production rates and limits of the first four comets discovered by the NEOWISE Reactivation mission. 

The infrared wavelengths provide a thermal emission and reflected light dust signal that can characterize a unique regime of dust particle sizes through analysis of the dust coma morphologies \citep{2015DPS....4750609K}. The NEOWISE multi-epoch observations of many of the comets detected so far provide characterization of long-term cometary behavior regarding these aspects of dust and gas emission. The gas and dust properties of the Reactivation Year 1 and 2 survey comet sample will be described in a later work.

\section{Methods}

\subsection{Extraction of Detections}

The methodology for extracting detections of minor planets from the NEOWISE source lists, as well as methods of diameter and albedo computation follows the description in \citet{Nugent15}, with the one exception described in Section 3.2. As was done in that work, the NEOWISE source lists were searched using the positions and times reported for each minor planet in the MPC's archival files NumObs.txt and UnnObs.txt\footnote{\url{http://www.minorplanetcenter.net/iau/ECS/MPCAT-OBS/MPCAT-OBS.html}}. NEOWISE reports detections to the MPC three times a week. By querying the MPC archive after the conclusion of Year 2 operations, we restrict our analysis to those detections of minor planets that were reported to and confirmed by the MPC.

These detections were converted into IRSA Catalog Query Engine format\footnote{\url{http://irsa.ipac.caltech.edu/applications/Gator/GatorAid/irsa/QuickGuidetoGator.htm}}, and were used to query the NEOWISE-R Single Exposure Level 1b (L1b) Source Table, which is served by the NASA/IPAC Infrared Science Archive (IRSA). The NEOWISE Reactivation data are described in detail in the NEOWISE Reactivation Explanatory Supplement \citep{2015Cutri}, which was updated in March 2016 to include single-exposure images and extracted source products from Reactivation Year 2. The Single Exposure (L1b) Source table was queried to find sources within 2 arcseconds of the reported position in the MPC files. For this query, detection time is constrained to be within two seconds of the reported time. The resulting table is a list of all sources corresponding to reported MPC detections from single exposures, with associated MPC designations for each detection.

Several steps are taken to prevent confusion of small body detections with fixed background sources such as stars and galaxies. We reference the WISE All-Sky Source Catalog, which is derived from a co-add of multiple exposures, covering the sky. This is a significantly deeper image than the individual L1b images, and pixel outlier rejection suppresses moving solar system objects. Therefore, it is useful for identifying fixed sources in the L1b images. The WISE Moving Object Pipeline System (WMOPS), which identifies moving objects in the NEOWISE images, compares single-exposure detections to reference images before any detections are submitted to the MPC. However, as an additional precaution, we also compare the single-exposure detection list to the All-Sky Catalog. Any single-exposure detections found to be within 6.5 arcseconds (the size of the 3.4 and 4.6 $\mu$m NEOWISE point-spread function) of a WISE All-Sky Source Catalog source with SNR $\ge 3$ were removed.

The resulting asteroid detection table was then stripped of measurements with associated poor quality flags. Each NEOWISE detection is graded for quality, as described in \citet{2015Cutri}. Detections with ``$ph\_qual$'' values of ``A'', ``B'', or ``C'' were accepted, this photometric quality grade ensures that the source was detected in the band with a flux signal-to-noise ratio $<2$. Additionally, detections must have ``$cc\_flags$'' values of ``0'' or ``p'', indicating that either the source was unaffected by known artifacts (``0''), or perhaps is impacted by a latent image left by a bright source (``p''). The value of ``p'' is conservative; it indicates the source is likely unaffected by a latent image, but possibly may be slightly contaminated. Finally, only frames graded ``$qual\_frame$'=``10'' or highest quality by the quality assurance process were used.

The WISE Science Data System pipeline profile-fitting magnitudes are used for each band \citep{2015Cutri}. A minimum of three detections with measurement uncertainties $\sigma_{mag} \le 0.25$ mag were required for thermal fits. Saturated detections, with a W1 magnitude $ \le 8.0$ or a W2 magnitude $ \le 7.0$, were discarded. The photometric measurements used for each asteroid are listed in Table \ref{tab:obs}.

\subsection{$H$ and $G$ values}

For each diameter, a corresponding albedo is also calculated, using an absolute visual magnitude $H$ and IAU phase slope parameter $G$ \citep{bowell89}. Therefore, the accuracy of albedos calculated from diameter measurements depends on the accuracy $H$ and $G$ values. The MPC provides $H$ and $G$ values as part of its catalog service; however, the default catalog values may be affected by various systematic effects \citep{2015Veres,Williams.2012a}. Known issues include values calculated from observations submitted with uncertain photometric calibrations and a bias towards discovering asteroids  when their longest axis faces Earth \citep{Jedicke2002}.

Corrected or newly-derived $H$ and $G$ values have been published by \citet{2009Warner, Pravec.2012a,Williams.2012a} and \citet{2015Veres}. The largest of these $H$ and $G$ datasets is from \citet{Williams.2012a} with $\sim 337000$ numbered asteroids, and \citet{2015Veres} with $\sim 250000$ objects observed by Pan-STARRS PS1. The \citet{Williams.2012a} dataset is slated to be incorporated into the MPC catalog (G. Williams, personal communication, May 2nd 2016). For this reason, and because it is more extensive, we used these corrected $H$ and $G$ values in this work when they were available for the asteroids in our sample. This is a departure from the methods in \citet{Nugent15}, which employed MPC database $H$ and $G$ values as no large replacement dataset was available at that time. Unless specified otherwise, $G$ is assumed to be  $0.15\pm0.1$ mag, and the error in H is assumed to be $\pm0.3$ mag.

\subsection{Diameter and albedo calculations}

The effective diameter $d$ of each asteroid and geometric optical albedo $p_v$ were then calculated from the resulting verified, high-quality minor planet measurements using the Near-Earth Asteroid Thermal Model \citep[NEATM;][]{Harris98}.  The implementation used in this work is detailed in  \citet{Mainzer2011b}.  It assumes a spherical object with no rotation, no nightside emission, and a temperature distribution given by:
\begin{equation}
T(\theta)=T_{max} \cos^{1/4}(\theta) \quad \textrm{for} \quad 0 \leq \theta \leq \pi/2
\label {eq:NEATM}
\end{equation}
where $\theta$  is the angular distance from the sub-solar point. $T_{max}$ is the sub-solar temperature, defined as:
\begin{equation}
T_{max}=\left( \frac{(1-A)S}{\eta \epsilon \sigma}\right)^{1/4}
\end{equation}
where $A$ is the bolometric Bond albedo, $S$ is the solar flux at the asteroid, $\eta$ is  the beaming parameter, $\epsilon$ is the emissivity, and $\sigma$ is the Stefan-Boltzmann constant. The beaming parameter $\eta$ adjusts the temperature distribution, and variation of $\eta$ can be due to non-spherical shapes, rotation rates, spin pole orientation with respect to observer, surface thermal inertia, phase effects, etc.

After a best-fit diameter is found, twenty-five Monte Carlo trials were run to evaluate the errors introduced by the uncertainty in the flux measurements. The corresponding uncertainties in diameter and albedo, along with the $H$ and $G$ values used as inputs to the thermal model, are reported in Tables \ref{tab:neoyr2} and \ref{tab:mbanew}. 

The NEOWISE survey cadence observes each object over $\sim1.5$ days on average, and sometimes re-observes an object $\sim3$ to $\sim6$ months later at a different distance and viewing geometry. These separate epochs, defined as observations separated by $>10$ days, the typical amount of time for viewing geometry of NEOs to change significantly, were fit separately. 

NEAs were treated differently than Mars-crossing and main belt asteroids, because of the different characteristics of the populations and different phase angles as demonstrated in \citet{Mainzer2011d}  and \citet{Masiero11}. In most cases, NEAs were fit with $\eta=1.4 \pm 0.5$. If both bands were thermally dominated, a beaming parameter was fit. A ratio of $p_{IR}/p_{V}= 1.6\pm  1.0$ was assumed for NEAs. Mars-crossing and main belt asteroids were fit with $\eta=0.95 \pm 0.2$, and the ratio of $p_{IR}/p_{V}$ was taken to be $1.5 \pm  0.1$ in most cases. These assumptions were necessary because if only one thermally dominated band is available, a beaming parameter cannot be fit; similarly, with only the 3.4 and 4.6 $\mu$m bands, we cannot fit $p_{IR}$ because there is not enough information to constrain it. 

As noted in Tables \ref{tab:neoyr2} and \ref{tab:mbanew}, some objects were fit with alternative beaming parameters and $p_{IR}/p_{V}$ ratios. In rare cases the standard assumption of $\eta=1.4 \pm 0.5$ and $\eta=0.95\pm0.2$ for NEAs and MBAs, respectively, lead to poor fits. Poor fits are indicated by $abs(H_{observed}-H_{modeled}) > 0.5$ or unphysical values of $p_V$, generally taken to be $p_V < \sim0.02$, $p_V > \sim0.6$ ).  In cases where a poor fit is obtained, we use the constraints on H magnitude errors and physical limits on albedo to exclude unphysical results, and rule out certain beaming and $p_{IR}/p_{V}$ values. A series of broadly-spaced beaming values (in increments of 0.2) and $p_{IR}/p_{V}$ ratios (in increments of 0.5) were tried; in these few cases, the associated errors were increased. These spacings were chosen so that in most cases only a single pair of beaming and  $p_{IR}/p_{V}$ ratios would produce a good fit. 

NEATM is only effective if at least one of the wavelength bands employed is dominated by thermal emission. Therefore, any object found to have $<75\%$ thermally emitted light (generally the cooler outer main belt objects) in both bands was removed from the results. This determination is made after an initial fit to the object is completed and estimates of thermally emitted and reflected light can be computed.

\section{Results}

Thermal fit results for NEAs are presented in Table \ref{tab:neoyr2}; Table \ref{tab:mbanew} contains the fit results for  Mars-crossing and main belt asteroids. When objects were observed at multiple epochs, a measurement of diameter and albedo is given for each epochs. 

\begin{figure}[h!]
  \caption{Comparison between asteroid diameters (top) and albedos (bottom) measured in this work with $H$ and $G$ values from the MPC (blue), diameters for the same objects measured in this work with revised $H$ and $G$ values from \citet{Williams.2012a} (black), and diameters for the same objects measured using previous NEOWISE measurements, which employed $H$ and $G$ values from the MPC (green). The bimodal structure of the albedo distribution is due to the populations of bright S-type ($p_V=0.25$) and dark C-type ($p_V=0.06$) objects in the main belt.}
  \centering
  \includegraphics[width=0.7\textwidth]{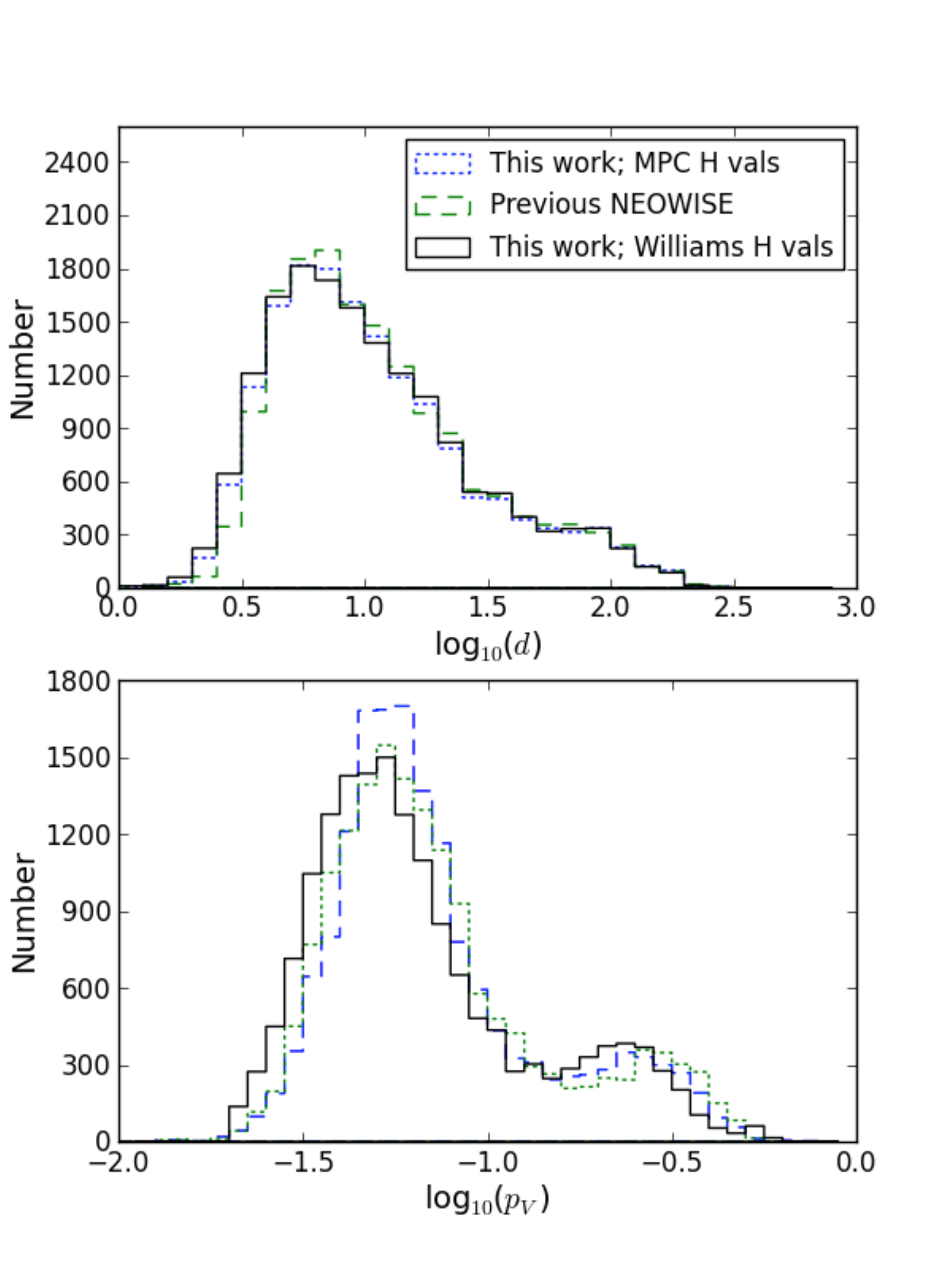}
  \label{fig:adcomp}
\end{figure}

Some asteroids have diameters and albedos calculated from earlier NEOWISE measurements \citep{Mainzer2011d, Masiero11,Mainzer12,Masiero12,Nugent15}. Figure \ref{fig:adcomp} is a histogram of the diameters and albedos for these objects measured from Reactivation Year 2 data and previous work. This figure also compares the results of the corrected $H$ and $G$ values from \citet{Williams.2012a}. Although distributions of diameter and albedo for this work are comparable to previous NEOWISE results, the incorporation of the revised $H$ and $G$ values does shift the albedo distribution towards slightly lower values. The implementation of the \citet{Williams.2012a} $H$ and $G$ values did not change the diameters of the ensemble of NEAs or other asteroids in a statistically significant way (see  Figure \ref{fig:adcomp}).

\begin{figure}[h!]
  \caption{Top: Comparison of diameters and albedos derived via radar, stellar occultations, and spacecraft flybys to the values calculated in this paper. The dashed red line shows a 1:1 relation. Bottom: Histograms of the fractional differences between the NEOWISE diameters ($\%\Delta d$, left) and albedos ($\%\Delta p_V$, right) and those derived from other methods. Dashed red line is best-fit Gaussian, with the fitted $\sigma$ given in the legends.}
  \centering
    \includegraphics[width=1.0\textwidth]{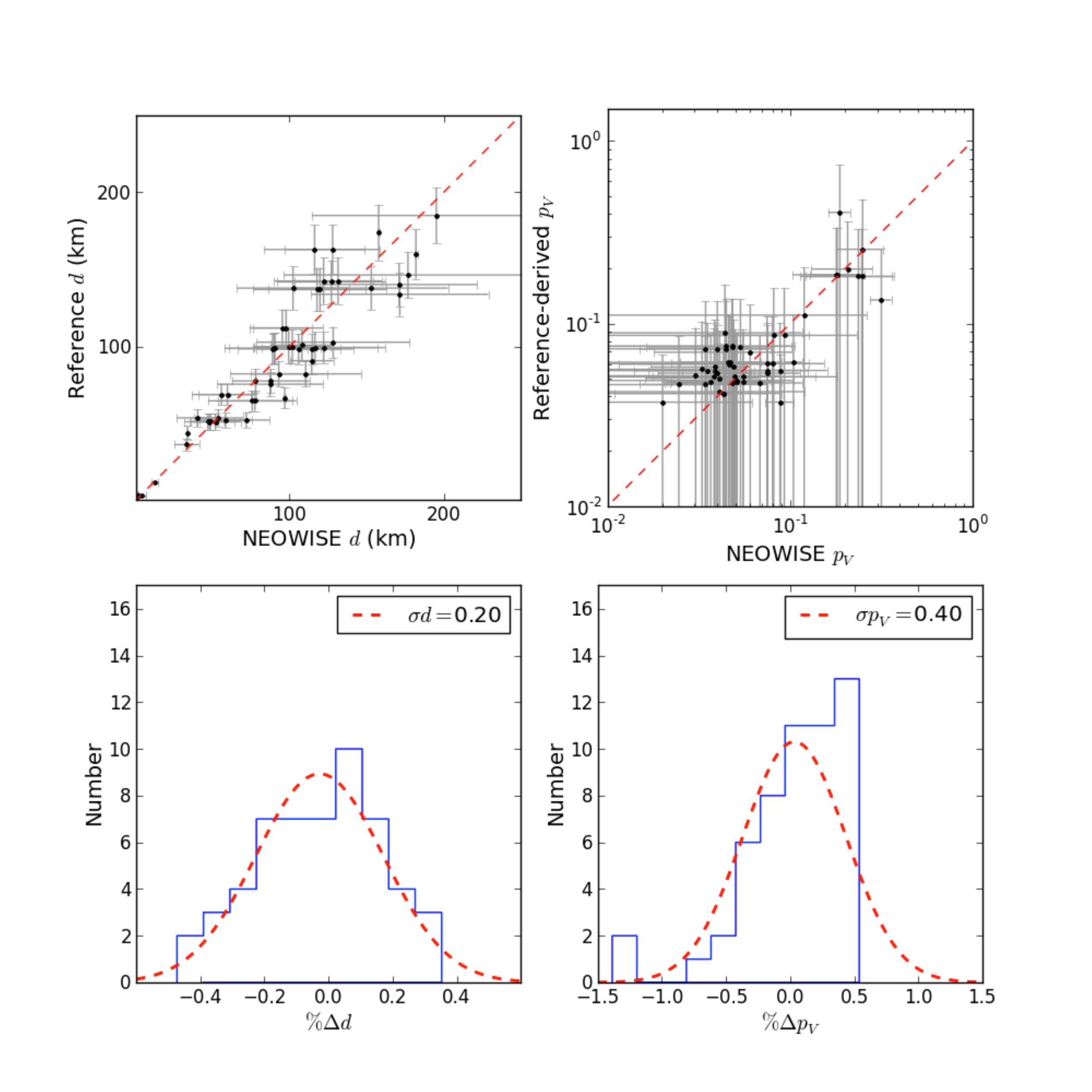}
    \label{fig:radar}
\end{figure}

When possible, diameters calculated from this work were compared to diameters calculated by independent methods (Figure \ref{fig:radar}). Twenty-three objects have diameters calculated via stellar occultations \citep{2006Shevchenko}, eleven have radar-derived shapes 
\citep{BennerAstIV_temp}, and two, (951) Gaspra and (253) Mathilde, were observed by spacecraft and had shape and size determined from resulting images \citep{Thomas94,1999Thomas}. These comparison cases were not preselected on light curve amplitude. When three-dimensional shapes were known, comparison was made to the average of the length of each axis. As illustrated in Figure \ref{fig:radar}, a Gaussian fit to a histogram of ($D_{NEOWISE} - D_{reference})/D_{NEOWISE}$ gives $\sigma = 20\%$, 
and a Gaussian fit to a histogram of $(p_{v-NEOWISE} - p_{v-reference})/p_{v-NEOWISE}$ gives $\sigma=40\%$. 

We report the Gaussian-fit $1-\sigma$ uncertainty of $20\%$ on diameter, and $40\%$ on albedo, based on the comparison to diameter measurements made with other techniques known to produce highly accurate diameters. This encompasses the systematic uncertainties in the comparison measurements (radar, stellar occultation, and spacecraft measurements), the range of ways that actual objects do not precisely match with the assumptions of NEATM, as well as the color corrections derived for the WISE filters \citep{Wright10WISE}. 

\begin{figure}[h!]
  \caption{Diameters and albedos from NEOWISE measurements of previously known NEAs (teal circles) and NEOWISE NEA discoveries (black squares) made during years 1 and 2 of the Reactivation. NEOWISE continues to detect large objects $>100$ m, and many discoveries are dark. Error bars on previously known objects were omitted for clarity.}
  \centering
  \includegraphics[width=0.7\textwidth]{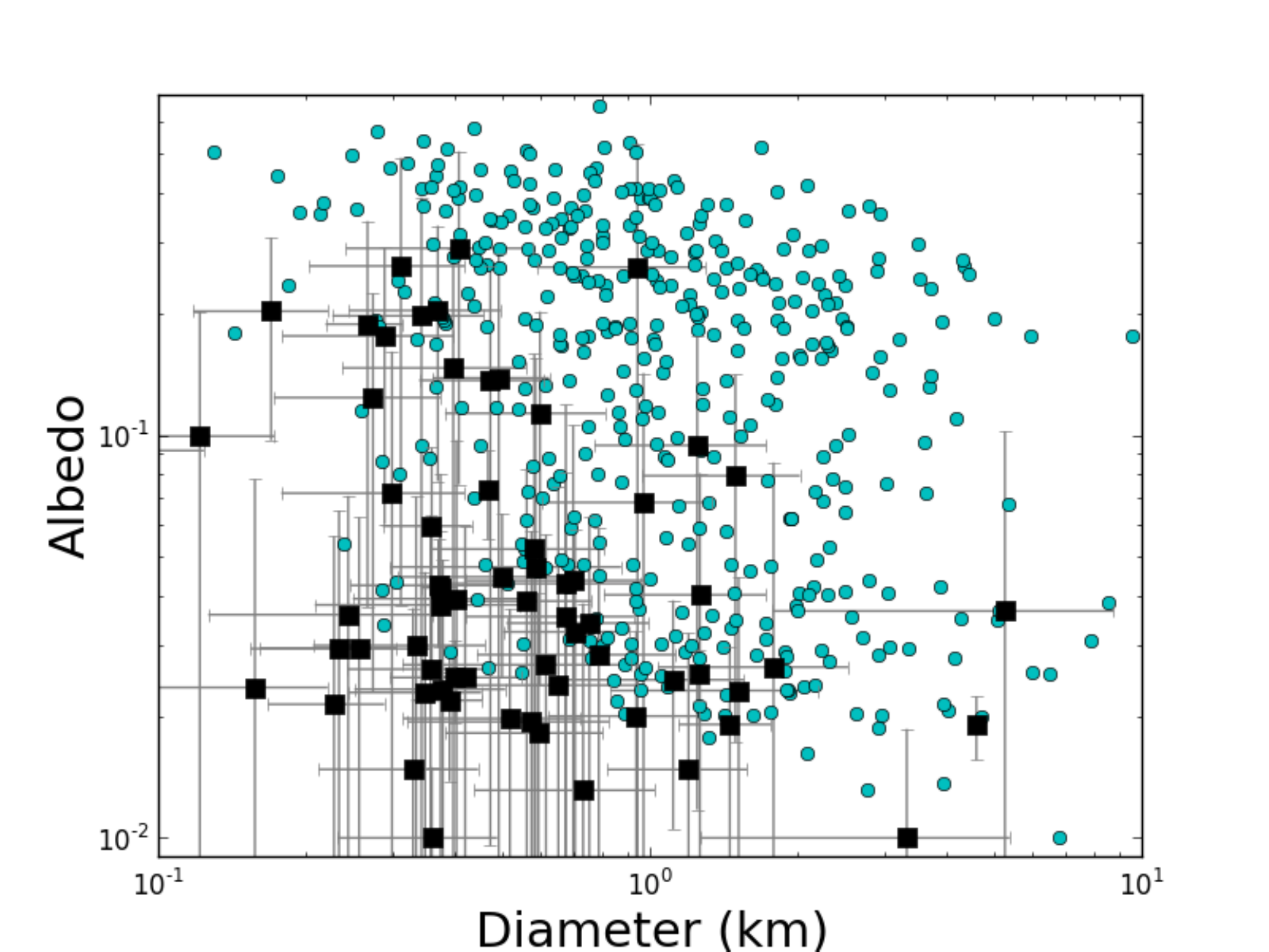}
  \label{fig:disc}
\end{figure}

The diameters and albedos of NEOWISE Reactivation discoveries are compared with the diameters and albedos of objects detected during Reactivation operations in Figure \ref{fig:disc}.  NEOWISE continues to discover large objects ($ > 100$ m), as well as low-albedo objects. 

\subsection{Potentially Hazardous Asteroids}

Potentially Hazardous Asteroids (PHAs) have been defined as objects with $H \le 22.0$ mag and a Minimum Orbit Intersection Distance (MOID) of 0.05 AU. The MOID is a measurement of the smallest distance between two orbits \citep{Sitarski68, Gronchi05}. Since many NEAs do not have measured diameters, the $H$ limit was used as a proxy for size. An object with $p_V= \sim 0.14$ and $H=22.0$ mag corresponds to an object $\sim140$ m in diameter.

Using the PHA definition as defined by $H$ limit, five NEOWISE Reactivation Year 2 discoveries are considered PHAs. However, eight NEOWISE Reactivation Year 2 discoveries are larger than 140 m in diameter and have a MOID $ \le0.05$ AU, and therefore should be classified as PHAs. With the availability of more diameter measurements of NEAs from NEOWISE, the Spitzer Space Telescope \citep{2010Trilling}, and ground-based facilities, sizes should be taken into consideration when designating PHAs as suggested in \citet{NEOWISE_PHA}. The fraction of PHAs within the NEOWISE NEA discoveries remains virtually constant across Year 1 and Year 2 of the Reactivation mission, and is nearly a factor of three higher than ground-based surveys\footnote{\url{http://neo.jpl.nasa.gov/stats/}}.

\subsection{NHATS}
The Near-Earth Object Human Space Flight Accessible Targets Study \citep[NHATS,][]{BarbeeNHATS} aims to identify the asteroids that would be most accessible to a crewed mission to an asteroid\footnote{\url{http://neo.jpl.nasa.gov/nhats/}}. NHATS-compliant targets must pass a series of restrictions, including Earth departure dates before 31 Dec 2040, total mission $\Delta V$ $\le 12$ km s$^{-1}$, and a minimum NEA stay time of 8 days. Many of these objects do not have measured physical properties. The NEOWISE Year 2 Reactivation mission measured diameters and albedos for eight objects (Table \ref{tab:NHATS}). Two, (35107) 1991 VH  and (363505) 2003 UC$_{20}$, were observed during Reactivation Year 1.

\section{Conclusion}
NEOWISE continues its mission to discover, track, and characterize minor planets. This release of diameters and albedos for 9,092 asteroids measured using NEOWISE Year 2 observations increases the total number of asteroids with measured diameters and albedos by 1,440, enabling further studies of NEAs and other asteroids by the scientific community and provides multi-epoch infrared observations that support more detailed thermophysical modeling studies. Comparison to diameters measured by other methods shows that measured diameters continue to be accurate to $\sim 20+\%$ during the Year 2 Reactivation mission. NEOWISE continues to preferentially discover large ($>100$m), low-albedo NEOs.

\section{Acknowledgments}

The authors thank the reviewer, Valerio Carruba, for his careful review that improved the quality of this manuscript.

This publication makes use of data products from the Wide-field Infrared Survey Explorer, which is a joint project of the University of California, Los Angeles, and JPL/California Institute of Technology, funded by NASA. This publication also makes use of data products from NEOWISE, which is a project of the JPL/California Institute of Technology, funded by the Planetary Science Division of NASA. The JPL High-Performance Computing Facility used for our simulations is supported by the JPL Office of the CIO.

This research has made use of the NASA/ IPAC Infrared Science Archive, which is operated by the Jet Propulsion Laboratory, California Institute of Technology, under contract with the National Aeronautics and Space Administration.

 This project used data obtained with the Dark Energy Camera (DECam), which was constructed by the Dark Energy Survey (DES) collaboration.
Funding for the DES Projects has been provided by 
the U.S. Department of Energy, 
the U.S. National Science Foundation, 
the Ministry of Science and Education of Spain, 
the Science and Technology Facilities Council of the United Kingdom, 
the Higher Education Funding Council for England, 
the National Center for Supercomputing Applications at the University of Illinois at Urbana-Champaign, 
the Kavli Institute of Cosmological Physics at the University of Chicago, 
the Center for Cosmology and Astro-Particle Physics at the Ohio State University, 
the Mitchell Institute for Fundamental Physics and Astronomy at Texas A\&M University, 
Financiadora de Estudos e Projetos, Funda{\c c}{\~a}o Carlos Chagas Filho de Amparo {\`a} Pesquisa do Estado do Rio de Janeiro, 
Conselho Nacional de Desenvolvimento Cient{\'i}fico e Tecnol{\'o}gico and the Minist{\'e}rio da Ci{\^e}ncia, Tecnologia e Inovac{\~a}o, 
the Deutsche Forschungsgemeinschaft, 
and the Collaborating Institutions in the Dark Energy Survey. 
The Collaborating Institutions are 
Argonne National Laboratory, 
the University of California at Santa Cruz, 
the University of Cambridge, 
Centro de Investigaciones En{\'e}rgeticas, Medioambientales y Tecnol{\'o}gicas-Madrid, 
the University of Chicago, 
University College London, 
the DES-Brazil Consortium, 
the University of Edinburgh, 
the Eidgen{\"o}ssische Technische Hoch\-schule (ETH) Z{\"u}rich, 
Fermi National Accelerator Laboratory, 
the University of Illinois at Urbana-Champaign, 
the Institut de Ci{\`e}ncies de l'Espai (IEEC/CSIC), 
the Institut de F{\'i}sica d'Altes Energies, 
Lawrence Berkeley National Laboratory, 
the Ludwig-Maximilians Universit{\"a}t M{\"u}nchen and the associated Excellence Cluster Universe, 
the University of Michigan, 
{the} National Optical Astronomy Observatory, 
the University of Nottingham, 
the Ohio State University, 
the University of Pennsylvania, 
the University of Portsmouth, 
SLAC National Accelerator Laboratory, 
Stanford University, 
the University of Sussex, 
and Texas A\&M University.

This work makes use of observations from the LCOGT network.

The authors wish to thank G. Williams, for providing the corrected $H$ and $G$ values from his dissertation used in this work.

This publication makes use of observations obtained at the Gemini Observatory, which is operated by the Association of Universities for Research in Astronomy, Inc., under a cooperative agreement with the NSF on behalf of the Gemini partnership: the National Science Foundation (United States), the National Research Council (Canada), CONICYT (Chile), Ministerio de Ciencia, Tecnolog\'{i}a e Innovaci\'{o}n Productiva (Argentina), and Minist\'{e}rio da Ci\^{e}ncia, Tecnologia e Inova\c{c}\~{a}o (Brazil). Observing Program IDs: GS-2015A-LP-3, GS-2015B-LP-3.

\bibliographystyle{apj_cn}

\clearpage

\begin{deluxetable}{rrrr}
\tabletypesize{\scriptsize}

\tablecaption{NEOWISE magnitudes for the NEAs modeled in this paper. Listed are MPC-packed name, the time of the observation in modified Julian date (MJD), and the magnitude in the $3.4\mu$m (W1) and $4.6\mu$m bands (W2). Non-detections at a particular wavelength represent $95\%$ confidence limits \citep{Cutri12}. Observations for the first two objects only are shown; the remainder are available in electronic format through the journal website.}
\tablewidth{0pt}
\tablehead{
	\colhead{Object} & \colhead{MJD} & \colhead{W1 (mag)}& \colhead{W2 (mag)} }                                                                                   
\startdata
   01580 & 57059.5362951 & 14.638 $\pm$ 0.074 & 11.338 $\pm$ 0.024 \\  
   01580 & 57059.7989937 & 14.540 $\pm$ 0.068 & 11.339 $\pm$ 0.024 \\  
   01580 & 57059.9304065 & 14.564 $\pm$ 0.076 & 11.274 $\pm$ 0.025 \\  
   01580 & 57060.0618195 & 14.525 $\pm$ 0.084 & 11.290 $\pm$ 0.028 \\  
   01580 & 57162.4156505 & 11.563 $\pm$ 0.017 & 8.478 $\pm$ 0.015 \\  
   01580 & 57162.4812294 & 11.507 $\pm$ 0.017 & 8.436 $\pm$ 0.012 \\  
   01580 & 57162.5468084 & 11.586 $\pm$ 0.018 & 8.497 $\pm$ 0.013 \\  
   01580 & 57162.6780938 & 11.527 $\pm$ 0.019 & 8.436 $\pm$ 0.014 \\  
   01580 & 57162.8092518 & 11.449 $\pm$ 0.017 & 8.388 $\pm$ 0.013 \\  
   01580 & 57163.2684322 & 11.724 $\pm$ 0.018 & 8.624 $\pm$ 0.015 \\  
   01580 & 57163.5308722 & 11.825 $\pm$ 0.021 & 8.695 $\pm$ 0.014 \\  
   01580 & 57163.5309995 & 11.836 $\pm$ 0.021 & 8.443 $\pm$ 0.014 \\  
   01580 & 57163.7933154 & 11.879 $\pm$ 0.019 & 8.803 $\pm$ 0.013 \\  
   01580 & 57163.9246007 & 11.808 $\pm$ 0.021 & 8.737 $\pm$ 0.013 \\  
   01580 & 57164.1213378 & 11.317 $\pm$ 0.017 & 8.222 $\pm$ 0.013 \\  
   01580 & 57164.3837815 & 11.406 $\pm$ 0.021 & 8.286 $\pm$ 0.015 \\  
   01580 & 57164.7772554 & 11.442 $\pm$ 0.021 & 8.373 $\pm$ 0.015 \\  
   01580 & 57164.8429617 & 11.566 $\pm$ 0.021 & 8.512 $\pm$ 0.013 \\  
   01580 & 57164.9741197 & 11.507 $\pm$ 0.019 & 8.381 $\pm$ 0.012 \\  
   01580 & 57165.1052777 & 11.417 $\pm$ 0.020 & 8.371 $\pm$ 0.014 \\  
   01580 & 57165.2364356 & 11.388 $\pm$ 0.026 & 8.330 $\pm$ 0.014 \\  
   01580 & 57165.2365629 & 11.432 $\pm$ 0.021 & 8.316 $\pm$ 0.012 \\  
   01580 & 57189.9632607 & 12.328 $\pm$ 0.023 & 9.224 $\pm$ 0.014 \\  
   01580 & 57190.0288402 & 12.292 $\pm$ 0.023 & 9.209 $\pm$ 0.013 \\  
   01580 & 57190.0944191 & 12.292 $\pm$ 0.026 & 9.204 $\pm$ 0.015 \\  
   01580 & 57190.1601254 & 12.022 $\pm$ 0.028 & 8.925 $\pm$ 0.012 \\  
   01580 & 57190.2257043 & 12.318 $\pm$ 0.028 & 9.243 $\pm$ 0.014 \\  
   01580 & 57190.2912832 & 12.307 $\pm$ 0.022 & 9.198 $\pm$ 0.014 \\  
   01580 & 57190.3568621 & 12.424 $\pm$ 0.025 & 9.268 $\pm$ 0.015 \\  
   01620 & 57259.6667067 & 14.743 $\pm$ 0.082 & 12.908 $\pm$ 0.061 \\  
   01620 & 57259.7979924 & 14.789 $\pm$ 0.104 & 12.701 $\pm$ 0.055 \\  
   01620 & 57259.9291508 & 15.247 $\pm$ 0.130 & 13.368 $\pm$ 0.099 \\  
   01620 & 57260.0603091 & 15.713 $\pm$ 0.177 & 13.865 $\pm$ 0.138 \\  
   01620 & 57260.2570466 & 15.336 $\pm$ 0.134 & 13.066 $\pm$ 0.071 \\  
   01620 & 57260.2571739 & 15.263 $\pm$ 0.125 & 13.355 $\pm$ 0.092 \\  
   01620 & 57260.3227531 & 14.641 $\pm$ 0.075 & 12.867 $\pm$ 0.067 \\  
   01620 & 57260.3883323 & 15.517 $\pm$ 0.150 & 13.546 $\pm$ 0.158 \\  
   01620 & 57260.4539115 & 14.675 $\pm$ 0.079 & 12.747 $\pm$ 0.067 \\  
   01620 & 57260.5850698 & 14.916 $\pm$ 0.091 & 13.053 $\pm$ 0.069 \\  
   01620 & 57260.7818073 & 14.934 $\pm$ 0.108 & 12.796 $\pm$ 0.060 \\  
   01620 & 57260.913093 & 15.390 $\pm$ 0.135 & 13.504 $\pm$ 0.114 \\  
   01620 & 57261.1754097 & 15.203 $\pm$ 0.115 & 13.108 $\pm$ 0.070 \\  
\enddata
\label{tab:obs}
\end{deluxetable}
\begin{deluxetable}{rrrrrrrrrrr}
\tabletypesize{\scriptsize}
\tablecaption{Measured diameters ($d$) and albedos ($p_V$) of near-Earth asteroids observed during the NEOWISE Year 2 mission. Asteroids may be identified by numbers, provisional designations, or via the MPC packed format. Magnitude $H$, slope parameter $G$, and beaming $\eta$ used are given. The numbers of observations used in the 3.4 $\mu$m ($n_{W1}$) and 4.6 $\mu$m ($n_{W2}$) wavelengths are also reported, along with the amplitude of the 4.6 $\mu$m light curve (W2 amp., in mag).}
\tablewidth{0pt}
\tablehead{
	\colhead{Object} & \colhead{Packed} & \colhead{$H$}& \colhead{$G$}& \colhead{$d$ (km)} & \colhead{$p_V$}   & \colhead{$\eta$} & \colhead{$p_{IR}/p_{V}$} & \colhead{W2 amp.} & \colhead{$n_{W1}$} & \colhead{$n_{W2}$} }                                                                                     
\startdata
 1580           & 01580 & 14.90 & 0.12 & 7.91 $\pm$ 0.08 & 0.03 $\pm$ 0.01 & 1.40 $\pm$ 0.00 & 2.31 $\pm$ 0.10 & 0.07 & 4 & 4 \\ 
 1580           & 01580 & 14.90 & 0.12 & 4.19 $\pm$ 0.06 & 0.11 $\pm$ 0.02 & 1.40 $\pm$ 0.00 & 1.03 $\pm$ 0.10 & 0.34 & 7 & 7 \\ 
 1580           & 01580 & 14.90 & 0.12 & 5.37 $\pm$ 0.04 & 0.07 $\pm$ 0.01 & 1.40 $\pm$ 0.00 & 1.61 $\pm$ 0.10 & 0.58 & 18 & 18 \\ 
 1620           & 01620 & 15.41 & 0.24 & 1.96 $\pm$ 0.06 & 0.32 $\pm$ 0.04 & 1.40 $\pm$ 0.00 & 1.78 $\pm$ 0.10 & 1.16 & 13 & 13 \\ 
 1685           & 01685 & 14.45 & 0.24 & 3.91 $\pm$ 0.08 & 0.19 $\pm$ 0.02 & 1.40 $\pm$ 0.00 & 1.68 $\pm$ 0.10 & 1.04 & 10 & 11 \\ 
 1980           & 01980 & 13.87 & 0.24 & 4.36 $\pm$ 0.10 & 0.26 $\pm$ 0.03 & 1.40 $\pm$ 0.00 & 1.83 $\pm$ 0.10 & 0.99 & 18 & 18 \\ 
 1980           & 01980 & 13.87 & 0.24 & 4.31 $\pm$ 0.18 & 0.27 $\pm$ 0.05 & 1.40 $\pm$ 0.00 & 1.94 $\pm$ 0.10 & 1.36 & 31 & 31 \\ 
 1980           & 01980 & 13.87 & 0.24 & 4.47 $\pm$ 0.14 & 0.25 $\pm$ 0.06 & 1.40 $\pm$ 0.00 & 1.97 $\pm$ 0.10 & 1.46 & 27 & 29 \\ 
 2062           & 02062 & 17.30 & 0.24 & 0.73 $\pm$ 0.03 & 0.39 $\pm$ 0.05 & 1.40 $\pm$ 0.00 & 1.46 $\pm$ 0.10 & 0.25 & 6 & 6 \\ 
 2063           & 02063 & 17.37 & 0.24 & 1.03 $\pm$ 0.03 & 0.19 $\pm$ 0.03 & 1.40 $\pm$ 0.00 & 0.93 $\pm$ 0.10 & 0.24 & 5 & 5 \\ 
 3691           & 03691 & 14.98 & 0.24 & 2.08 $\pm$ 0.09 & 0.42 $\pm$ 0.11 & 1.40 $\pm$ 0.00 & 1.19 $\pm$ 0.10 & 0.73 & 21 & 21 \\ 
 4055           & 04055 & 14.99 & 0.43 & 3.21 $\pm$ 0.16 & 0.17 $\pm$ 0.03 & 1.40 $\pm$ 0.00 & 1.70 $\pm$ 0.10 & 0.91 & 6 & 7 \\ 
 4183           & 04183 & 14.35 & 0.24 & 3.73 $\pm$ 0.15 & 0.23 $\pm$ 0.04 & 1.40 $\pm$ 0.00 & 1.21 $\pm$ 0.10 & 0.33 & 9 & 9 \\ 
 5646           & 05646 & 15.45 & 0.24 & 2.45 $\pm$ 0.06 & 0.19 $\pm$ 0.03 & 1.40 $\pm$ 0.00 & 1.32 $\pm$ 0.10 & 0.36 & 39 & 39 \\ 
 5646           & 05646 & 15.45 & 0.24 & 2.50 $\pm$ 0.05 & 0.19 $\pm$ 0.03 & 1.40 $\pm$ 0.00 & 1.28 $\pm$ 0.10 & 0.66 & 74 & 78 \\ 
 5646           & 05646 & 15.45 & 0.24 & 2.51 $\pm$ 0.05 & 0.18 $\pm$ 0.03 & 1.40 $\pm$ 0.00 & 1.41 $\pm$ 0.10 & 0.35 & 46 & 47 \\ 
 5731           & 05731 & 15.53 & 0.12 & 6.51 $\pm$ 3.14 & 0.03 $\pm$ 0.03 & 1.00 $\pm$ 0.51 & 1.60 $\pm$ 0.10 & 0.33 & 0 & 8 \\ 
 5828           & 05828 & 16.30 & 0.24 & 1.43 $\pm$ 0.06 & 0.26 $\pm$ 0.05 & 1.40 $\pm$ 0.00 & 1.29 $\pm$ 0.10 & 0.38 & 6 & 6 \\ 
 7335           & 07335 & 17.82 & 0.24 & 0.73 $\pm$ 0.02 & 0.25 $\pm$ 0.04 & 1.40 $\pm$ 0.00 & 1.52 $\pm$ 0.10 & 0.78 & 8 & 8 \\ 
 7350           & 07350 & 17.21 & 0.24 & 1.92 $\pm$ 0.03 & 0.06 $\pm$ 0.01 & 1.40 $\pm$ 0.00 & 1.06 $\pm$ 0.10 & 0.12 & 4 & 5 \\ 
 7889           & 07889 & 15.31 & 0.24 & 1.82 $\pm$ 0.08 & 0.40 $\pm$ 0.04 & 1.40 $\pm$ 0.00 & 1.27 $\pm$ 0.10 & 0.77 & 13 & 13 \\ 
 9202           & 09202 & 16.16 & 0.24 & 1.51 $\pm$ 0.05 & 0.27 $\pm$ 0.04 & 1.40 $\pm$ 0.00 & 1.29 $\pm$ 0.10 & 0.18 & 5 & 6 \\ 
 9400           & 09400 & 14.98 & 0.24 & 3.69 $\pm$ 0.05 & 0.13 $\pm$ 0.02 & 1.40 $\pm$ 0.00 & 2.13 $\pm$ 0.10 & 0.20 & 10 & 10 \\ 
 11066           & 11066 & 15.36 & 0.24 & 2.10 $\pm$ 0.09 & 0.29 $\pm$ 0.04 & 1.40 $\pm$ 0.00 & 1.83 $\pm$ 0.10 & 0.49 & 5 & 5 \\ 
 11405           & 11405 & 15.37 & 0.24 & 3.62 $\pm$ 0.05 & 0.10 $\pm$ 0.02 & 1.40 $\pm$ 0.00 & 2.67 $\pm$ 0.10 & 0.25 & 13 & 13 \\ 
 21088           & 21088 & 14.47 & 0.24 & 2.79 $\pm$ 0.10 & 0.37 $\pm$ 0.06 & 1.40 $\pm$ 0.00 & 1.63 $\pm$ 0.10 & 0.45 & 12 & 12 \\ 
 22099           & 22099 & 18.19 & 0.24 & 0.52 $\pm$ 0.11 & 0.35 $\pm$ 0.22 & 1.40 $\pm$ 0.40 & 1.60 $\pm$ 0.10 & 0.81 & 0 & 4 \\ 
 23606           & 23606 & 18.37 & 0.24 & 0.87 $\pm$ 0.01 & 0.11 $\pm$ 0.02 & 1.40 $\pm$ 0.00 & 2.46 $\pm$ 0.10 & 1.67 & 60 & 60 \\ 
 26817           & 26817 & 19.11 & 0.24 & 1.17 $\pm$ 0.53 & 0.03 $\pm$ 0.08 & 1.40 $\pm$ 0.52 & 1.60 $\pm$ 0.10 & 0.41 & 0 & 6 \\ 
 35107           & 35107 & 17.02 & 0.24 & 0.91 $\pm$ 0.03 & 0.33 $\pm$ 0.04 & 1.40 $\pm$ 0.00 & 1.41 $\pm$ 0.10 & 0.93 & 27 & 27 \\ 
 38086           & 38086 & 17.63 & 0.24 & 0.64 $\pm$ 0.19 & 0.39 $\pm$ 0.25 & 1.40 $\pm$ 0.50 & 1.60 $\pm$ 0.10 & 0.71 & 0 & 6 \\ 
 38091           & 38091 & 16.61 & 0.24 & 2.49 $\pm$ 0.03 & 0.06 $\pm$ 0.01 & 1.40 $\pm$ 0.00 & 2.30 $\pm$ 0.10 & 0.97 & 12 & 13 \\ 
 52750           & 52750 & 16.74 & 0.24 & 0.96 $\pm$ 0.04 & 0.39 $\pm$ 0.06 & 1.40 $\pm$ 0.00 & 1.55 $\pm$ 0.10 & 0.73 & 22 & 24 \\ 
 52750           & 52750 & 16.74 & 0.24 & 0.93 $\pm$ 0.32 & 0.41 $\pm$ 0.25 & 1.40 $\pm$ 0.51 & 1.60 $\pm$ 0.10 & 0.19 & 0 & 5 \\ 
 53430           & 53430 & 16.85 & 0.24 & 1.34 $\pm$ 0.45 & 0.18 $\pm$ 0.17 & 1.40 $\pm$ 0.44 & 1.60 $\pm$ 0.10 & 0.78 & 0 & 10 \\ 
 68216           & 68216 & 16.66 & 0.24 & 0.99 $\pm$ 0.04 & 0.39 $\pm$ 0.06 & 1.40 $\pm$ 0.00 & 1.30 $\pm$ 0.10 & 0.38 & 8 & 8 \\ 
 68278           & 68278 & 18.50 & 0.12 & 1.46 $\pm$ 0.53 & 0.03 $\pm$ 0.03 & 1.40 $\pm$ 0.40 & 1.60 $\pm$ 0.10 & 0.52 & 0 & 23 \\ 
 85275           & 85275 & 16.45 & 0.24 & 2.50 $\pm$ 0.99 & 0.07 $\pm$ 0.06 & 1.40 $\pm$ 0.42 & 1.60 $\pm$ 0.10 & 0.53 & 0 & 14 \\ 
 85713           & 85713 & 16.01 & 0.24 & 3.03 $\pm$ 1.49 & 0.08 $\pm$ 0.12 & 1.40 $\pm$ 0.50 & 1.60 $\pm$ 0.10 & 0.37 & 0 & 6 \\ 
 85804           & 85804 & 15.47 & 0.24 & 2.27 $\pm$ 0.06 & 0.22 $\pm$ 0.04 & 1.40 $\pm$ 0.00 & 1.24 $\pm$ 0.10 & 0.17 & 11 & 11 \\ 
 85804           & 85804 & 15.47 & 0.24 & 2.19 $\pm$ 0.04 & 0.24 $\pm$ 0.02 & 1.40 $\pm$ 0.00 & 1.29 $\pm$ 0.10 & 0.21 & 13 & 14 \\ 
 85804           & 85804 & 15.47 & 0.24 & 2.84 $\pm$ 0.04 & 0.14 $\pm$ 0.02 & 1.40 $\pm$ 0.00 & 1.94 $\pm$ 0.10 & 0.58 & 21 & 22 \\ 
 85989           & 85989 & 17.03 & 0.24 & 1.60 $\pm$ 0.59 & 0.11 $\pm$ 0.17 & 1.40 $\pm$ 0.45 & 1.60 $\pm$ 0.10 & 0.39 & 0 & 4 \\ 
 86067           & 86067 & 16.52 & 0.24 & 1.50 $\pm$ 0.49 & 0.19 $\pm$ 0.14 & 1.40 $\pm$ 0.42 & 1.60 $\pm$ 0.10 & 0.54 & 0 & 11 \\ 
 86667           & 86667 & 17.59 & 0.24 & 0.74 $\pm$ 0.02 & 0.29 $\pm$ 0.05 & 1.40 $\pm$ 0.00 & 1.60 $\pm$ 0.10 & 0.76 & 4 & 4 \\ 
 86829           & 86829 & 16.11 & 0.24 & 1.81 $\pm$ 0.05 & 0.19 $\pm$ 0.03 & 1.40 $\pm$ 0.00 & 1.38 $\pm$ 0.10 & 0.13 & 6 & 6 \\ 
 88263           & 88263 & 15.73 & 0.24 & 5.10 $\pm$ 1.86 & 0.03 $\pm$ 0.04 & 1.40 $\pm$ 0.38 & 1.60 $\pm$ 0.10 & 0.93 & 0 & 18 \\ 
 88710           & 88710 & 18.14 & 0.34 & 0.75 $\pm$ 0.29 & 0.18 $\pm$ 0.14 & 1.40 $\pm$ 0.52 & 1.60 $\pm$ 0.10 & 0.99 & 0 & 71 \\ 
 90367           & 90367 & 18.11 & 0.24 & 2.06 $\pm$ 1.14 & 0.02 $\pm$ 0.07 & 1.40 $\pm$ 0.59 & 1.60 $\pm$ 0.10 & 0.48 & 0 & 13 \\ 
 90403           & 90403 & 17.78 & 0.24 & 0.57 $\pm$ 0.17 & 0.42 $\pm$ 0.24 & 1.40 $\pm$ 0.49 & 1.60 $\pm$ 0.10 & 0.49 & 0 & 5 \\ 
 90416           & 90416 & 18.58 & 0.24 & 0.98 $\pm$ 0.02 & 0.07 $\pm$ 0.01 & 1.40 $\pm$ 0.00 & 1.34 $\pm$ 0.10 & 0.07 & 5 & 5 \\ 
 100756           & A0756 & 16.48 & 0.24 & 1.81 $\pm$ 0.04 & 0.14 $\pm$ 0.02 & 1.40 $\pm$ 0.00 & 1.80 $\pm$ 0.10 & 1.16 & 15 & 15 \\ 
 105140           & A5140 & 15.81 & 0.24 & 1.97 $\pm$ 0.05 & 0.22 $\pm$ 0.04 & 1.40 $\pm$ 0.00 & 1.30 $\pm$ 0.10 & 0.58 & 8 & 9 \\ 
 108519           & A8519 & 17.94 & 0.24 & 1.43 $\pm$ 0.54 & 0.06 $\pm$ 0.09 & 1.40 $\pm$ 0.46 & 1.60 $\pm$ 0.10 & 0.52 & 0 & 12 \\ 
 112985           & B2985 & 15.66 & 0.24 & 3.65 $\pm$ 0.04 & 0.07 $\pm$ 0.01 & 1.40 $\pm$ 0.00 & 2.46 $\pm$ 0.10 & 0.12 & 11 & 11 \\ 
 112985           & B2985 & 15.66 & 0.24 & 5.12 $\pm$ 0.03 & 0.04 $\pm$ 0.01 & 1.40 $\pm$ 0.00 & 2.80 $\pm$ 0.10 & 0.15 & 18 & 18 \\ 
 137084           & D7084 & 16.52 & 0.24 & 1.23 $\pm$ 0.04 & 0.29 $\pm$ 0.04 & 1.40 $\pm$ 0.00 & 1.10 $\pm$ 0.10 & 0.48 & 4 & 4 \\ 
 137805           & D7805 & 16.77 & 0.24 & 2.24 $\pm$ 0.03 & 0.07 $\pm$ 0.01 & 1.40 $\pm$ 0.00 & 1.40 $\pm$ 0.10 & 0.31 & 17 & 17 \\ 
 137925           & D7925 & 16.25 & 0.24 & 1.36 $\pm$ 0.04 & 0.30 $\pm$ 0.05 & 1.40 $\pm$ 0.00 & 1.27 $\pm$ 0.10 & 0.29 & 4 & 5 \\ 
 140288           & E0288 & 16.84 & 0.24 & 1.26 $\pm$ 0.03 & 0.20 $\pm$ 0.04 & 1.40 $\pm$ 0.00 & 1.03 $\pm$ 0.10 & 0.34 & 8 & 8 \\ 
 140288           & E0288 & 16.84 & 0.24 & 1.21 $\pm$ 0.45 & 0.22 $\pm$ 0.23 & 1.40 $\pm$ 0.50 & 1.60 $\pm$ 0.10 & 0.74 & 0 & 5 \\ 
 141484           & E1484 & 16.64 & 0.24 & 1.00 $\pm$ 0.04 & 0.39 $\pm$ 0.05 & 1.40 $\pm$ 0.00 & 1.44 $\pm$ 0.10 & 0.26 & 8 & 9 \\ 
 141484           & E1484 & 16.64 & 0.24 & 1.02 $\pm$ 0.03 & 0.37 $\pm$ 0.04 & 1.40 $\pm$ 0.00 & 1.59 $\pm$ 0.10 & 0.39 & 16 & 16 \\ 
 142040           & E2040 & 16.31 & 0.24 & 1.26 $\pm$ 0.04 & 0.34 $\pm$ 0.04 & 1.40 $\pm$ 0.00 & 1.71 $\pm$ 0.10 & 0.36 & 43 & 44 \\ 
 152679           & F2679 & 16.43 & 0.12 & 4.18 $\pm$ 0.01 & 0.03 $\pm$ 0.00 & 1.40 $\pm$ 0.00 & 7.21 $\pm$ 0.10 & 1.22 & 29 & 29 \\ 
 152978           & F2978 & 19.73 & 0.24 & 0.32 $\pm$ 0.07 & 0.23 $\pm$ 0.13 & 1.40 $\pm$ 0.32 & 1.60 $\pm$ 0.10 & 0.74 & 0 & 8 \\ 
 152978           & F2978 & 19.73 & 0.24 & 0.37 $\pm$ 0.14 & 0.17 $\pm$ 0.14 & 1.40 $\pm$ 0.47 & 1.60 $\pm$ 0.10 & 0.57 & 0 & 5 \\ 
 153195           & F3195 & 17.94 & 0.24 & 1.32 $\pm$ 0.55 & 0.07 $\pm$ 0.07 & 1.40 $\pm$ 0.47 & 1.60 $\pm$ 0.10 & 0.77 & 0 & 14 \\ 
 153195           & F3195 & 17.94 & 0.24 & 1.60 $\pm$ 0.59 & 0.05 $\pm$ 0.04 & 1.40 $\pm$ 0.39 & 1.60 $\pm$ 0.10 & 0.45 & 0 & 7 \\ 
 154807           & F4807 & 18.72 & 0.24 & 0.47 $\pm$ 0.01 & 0.26 $\pm$ 0.03 & 1.40 $\pm$ 0.00 & 1.69 $\pm$ 0.10 & 0.92 & 14 & 15 \\ 
 155110           & F5110 & 17.66 & 0.24 & 0.68 $\pm$ 0.03 & 0.33 $\pm$ 0.04 & 1.40 $\pm$ 0.00 & 1.87 $\pm$ 0.10 & 0.34 & 6 & 6 \\ 
 155110           & F5110 & 17.66 & 0.24 & 0.75 $\pm$ 0.32 & 0.27 $\pm$ 0.23 & 1.40 $\pm$ 0.54 & 1.60 $\pm$ 0.10 & 0.37 & 0 & 5 \\ 
 159459           & F9459 & 15.98 & 0.24 & 1.83 $\pm$ 0.09 & 0.21 $\pm$ 0.03 & 1.40 $\pm$ 0.00 & 1.11 $\pm$ 0.10 & 0.38 & 5 & 6 \\ 
 159504           & F9504 & 16.99 & 0.05 & 2.31 $\pm$ 0.02 & 0.05 $\pm$ 0.01 & 1.40 $\pm$ 0.00 & 1.73 $\pm$ 0.10 & 0.24 & 9 & 9 \\ 
 159686           & F9686 & 16.65 & 0.24 & 1.80 $\pm$ 0.03 & 0.12 $\pm$ 0.02 & 1.40 $\pm$ 0.00 & 3.56 $\pm$ 0.10 & 0.28 & 8 & 9 \\ 
 159929           & F9929 & 17.75 & 0.24 & 2.62 $\pm$ 1.20 & 0.02 $\pm$ 0.08 & 1.40 $\pm$ 0.45 & 1.60 $\pm$ 0.10 & 0.34 & 0 & 8 \\ 
 161989           & G1989 & 17.43 & 0.24 & 0.64 $\pm$ 0.02 & 0.46 $\pm$ 0.09 & 1.40 $\pm$ 0.00 & 1.81 $\pm$ 0.10 & 1.16 & 64 & 69 \\ 
 162080           & G2080 & 19.89 & 0.24 & 0.68 $\pm$ 0.32 & 0.04 $\pm$ 0.05 & 1.40 $\pm$ 0.55 & 1.60 $\pm$ 0.10 & 0.94 & 0 & 20 \\ 
 162463           & G2463 & 17.99 & 0.24 & 0.93 $\pm$ 0.35 & 0.13 $\pm$ 0.21 & 1.40 $\pm$ 0.49 & 1.60 $\pm$ 0.10 & 0.85 & 0 & 21 \\ 
 162463           & G2463 & 17.99 & 0.24 & 0.98 $\pm$ 0.35 & 0.12 $\pm$ 0.14 & 1.40 $\pm$ 0.43 & 1.60 $\pm$ 0.10 & 0.40 & 0 & 6 \\ 
 162567           & G2567 & 20.18 & 0.24 & 0.29 $\pm$ 0.12 & 0.18 $\pm$ 0.16 & 1.40 $\pm$ 0.56 & 1.60 $\pm$ 0.10 & 0.19 & 0 & 5 \\ 
 163760           & G3760 & 16.53 & 0.24 & 2.35 $\pm$ 0.77 & 0.08 $\pm$ 0.08 & 1.40 $\pm$ 0.39 & 1.60 $\pm$ 0.10 & 0.40 & 0 & 9 \\ 
 163899           & G3899 & 17.36 & 0.24 & 0.80 $\pm$ 0.02 & 0.31 $\pm$ 0.04 & 1.40 $\pm$ 0.00 & 1.35 $\pm$ 0.10 & 1.97 & 24 & 24 \\ 
 164206           & G4206 & 17.86 & 0.24 & 1.13 $\pm$ 0.55 & 0.10 $\pm$ 0.09 & 1.40 $\pm$ 0.55 & 1.60 $\pm$ 0.10 & 0.94 & 0 & 8 \\ 
 172034           & H2034 & 17.67 & 0.24 & 0.66 $\pm$ 0.17 & 0.34 $\pm$ 0.25 & 1.40 $\pm$ 0.44 & 1.60 $\pm$ 0.10 & 1.17 & 0 & 24 \\ 
 173689           & H3689 & 18.28 & 0.24 & 0.73 $\pm$ 0.29 & 0.16 $\pm$ 0.19 & 1.40 $\pm$ 0.55 & 1.60 $\pm$ 0.10 & 0.17 & 0 & 5 \\ 
 190161           & J0161 & 16.67 & 0.24 & 3.05 $\pm$ 0.02 & 0.04 $\pm$ 0.01 & 1.40 $\pm$ 0.00 & 1.79 $\pm$ 0.10 & 0.36 & 28 & 30 \\ 
 200754           & K0754 & 18.67 & 0.24 & 0.56 $\pm$ 0.21 & 0.19 $\pm$ 0.15 & 1.40 $\pm$ 0.52 & 1.60 $\pm$ 0.10 & 0.29 & 0 & 5 \\ 
 206378           & K6378 & 18.68 & 0.06 & 0.37 $\pm$ 0.02 & 0.44 $\pm$ 0.19 & 1.40 $\pm$ 0.00 & 1.85 $\pm$ 0.10 & 0.75 & 20 & 20 \\ 
 212359           & L2359 & 16.98 & 0.24 & 1.25 $\pm$ 0.40 & 0.18 $\pm$ 0.18 & 1.40 $\pm$ 0.42 & 1.60 $\pm$ 0.10 & 0.33 & 0 & 9 \\ 
 237805           & N7805 & 17.63 & 0.24 & 0.69 $\pm$ 0.26 & 0.33 $\pm$ 0.22 & 1.40 $\pm$ 0.57 & 1.60 $\pm$ 0.10 & 1.47 & 0 & 45 \\ 
 241662           & O1662 & 17.64 & 0.24 & 0.91 $\pm$ 0.02 & 0.19 $\pm$ 0.03 & 1.40 $\pm$ 0.00 & 1.56 $\pm$ 0.10 & 0.38 & 10 & 10 \\ 
 241662           & O1662 & 17.64 & 0.24 & 0.81 $\pm$ 0.26 & 0.24 $\pm$ 0.22 & 1.40 $\pm$ 0.47 & 1.60 $\pm$ 0.10 & 0.44 & 0 & 8 \\ 
 248590           & O8590 & 16.82 & 0.24 & 3.35 $\pm$ 1.04 & 0.03 $\pm$ 0.03 & 0.80 $\pm$ 0.40 & 1.50 $\pm$ 0.10 & 0.27 & 0 & 13 \\ 
 256412           & P6412 & 17.17 & 0.24 & 2.90 $\pm$ 0.02 & 0.03 $\pm$ 0.01 & 1.00 $\pm$ 0.20 & 4.06 $\pm$ 0.10 & 0.57 & 15 & 15 \\ 
 275611           & R5611 & 18.24 & 0.24 & 1.48 $\pm$ 0.01 & 0.04 $\pm$ 0.01 & 1.40 $\pm$ 0.00 & 2.67 $\pm$ 0.10 & 0.70 & 24 & 25 \\ 
 276049           & R6049 & 16.50 & 0.12 & 2.24 $\pm$ 0.02 & 0.09 $\pm$ 0.02 & 1.00 $\pm$ 0.20 & 0.23 $\pm$ 0.10 & 0.24 & 11 & 11 \\ 
 276049           & R6049 & 16.50 & 0.12 & 4.71 $\pm$ 2.84 & 0.02 $\pm$ 0.04 & 1.00 $\pm$ 0.60 & 1.60 $\pm$ 0.10 & 0.37 & 0 & 12 \\ 
 276786           & R6786 & 18.11 & 0.24 & 1.72 $\pm$ 0.68 & 0.03 $\pm$ 0.07 & 1.40 $\pm$ 0.45 & 1.60 $\pm$ 0.10 & 0.85 & 0 & 12 \\ 
 285331           & S5331 & 18.47 & 0.24 & 0.66 $\pm$ 0.01 & 0.17 $\pm$ 0.02 & 1.40 $\pm$ 0.00 & 2.04 $\pm$ 0.10 & 0.32 & 6 & 6 \\ 
 285331           & S5331 & 18.47 & 0.24 & 0.65 $\pm$ 0.01 & 0.17 $\pm$ 0.02 & 1.40 $\pm$ 0.00 & 1.96 $\pm$ 0.10 & 0.65 & 10 & 10 \\ 
 294739           & T4739 & 17.39 & 0.24 & 0.74 $\pm$ 0.21 & 0.36 $\pm$ 0.19 & 1.40 $\pm$ 0.46 & 1.60 $\pm$ 0.10 & 0.35 & 0 & 7 \\ 
 297274           & T7274 & 16.90 & 0.24 & 1.21 $\pm$ 0.03 & 0.21 $\pm$ 0.03 & 1.40 $\pm$ 0.00 & 1.01 $\pm$ 0.10 & 0.64 & 5 & 5 \\ 
 303450           & U3450 & 20.86 & 0.24 & 0.18 $\pm$ 0.06 & 0.24 $\pm$ 0.12 & 1.40 $\pm$ 0.43 & 1.60 $\pm$ 0.10 & 0.62 & 0 & 7 \\ 
 307493           & U7493 & 18.95 & 0.24 & 1.43 $\pm$ 0.09 & 0.02 $\pm$ 0.00 & 1.22 $\pm$ 0.06 & 1.60 $\pm$ 0.10 & 0.34 & 10 & 10 \\ 
 311554           & V1554 & 18.80 & 0.24 & 0.38 $\pm$ 0.02 & 0.36 $\pm$ 0.05 & 1.40 $\pm$ 0.00 & 2.07 $\pm$ 0.10 & 0.72 & 7 & 8 \\ 
 326388           & W6388 & 18.26 & 0.24 & 1.15 $\pm$ 0.01 & 0.07 $\pm$ 0.01 & 1.40 $\pm$ 0.00 & 0.70 $\pm$ 0.10 & 0.39 & 11 & 12 \\ 
 337248           & X7248 & 20.00 & 0.15 & 0.85 $\pm$ 0.06 & 0.02 $\pm$ 0.01 & 1.40 $\pm$ 0.09 & 4.79 $\pm$ 0.10 & 1.01 & 6 & 6 \\ 
 337248           & X7248 & 20.00 & 0.15 & 0.61 $\pm$ 0.26 & 0.05 $\pm$ 0.06 & 1.40 $\pm$ 0.49 & 1.60 $\pm$ 0.10 & 0.34 & 0 & 7 \\ 
 345646           & Y5646 & 19.90 & 0.15 & 0.41 $\pm$ 0.01 & 0.12 $\pm$ 0.02 & 1.40 $\pm$ 0.00 & 8.52 $\pm$ 0.10 & 0.42 & 6 & 7 \\ 
 355770           & Z5770 & 18.40 & 0.15 & 1.20 $\pm$ 0.49 & 0.05 $\pm$ 0.08 & 1.40 $\pm$ 0.48 & 1.60 $\pm$ 0.10 & 1.68 & 0 & 57 \\ 
 363027           & a3027 & 19.50 & 0.15 & 0.58 $\pm$ 0.27 & 0.08 $\pm$ 0.11 & 1.40 $\pm$ 0.54 & 1.60 $\pm$ 0.10 & 0.30 & 0 & 4 \\ 
 363027           & a3027 & 19.50 & 0.15 & 0.69 $\pm$ 0.25 & 0.06 $\pm$ 0.07 & 1.40 $\pm$ 0.46 & 1.60 $\pm$ 0.10 & 0.19 & 0 & 6 \\ 
 363505           & a3505 & 18.10 & 0.15 & 1.88 $\pm$ 0.01 & 0.03 $\pm$ 0.00 & 1.40 $\pm$ 0.00 & 2.63 $\pm$ 0.10 & 0.70 & 26 & 28 \\ 
 373135           & b3135 & 19.50 & 0.15 & 1.05 $\pm$ 0.36 & 0.03 $\pm$ 0.01 & 1.40 $\pm$ 0.38 & 1.60 $\pm$ 0.10 & 0.88 & 0 & 4 \\ 
 381906           & c1906 & 17.90 & 0.15 & 0.52 $\pm$ 0.10 & 0.45 $\pm$ 0.27 & 1.40 $\pm$ 0.39 & 1.60 $\pm$ 0.10 & 0.43 & 0 & 5 \\ 
 385186           & c5186 & 17.70 & 0.15 & 0.81 $\pm$ 0.02 & 0.22 $\pm$ 0.03 & 1.40 $\pm$ 0.00 & 2.04 $\pm$ 0.10 & 0.57 & 9 & 9 \\ 
 385186           & c5186 & 17.70 & 0.15 & 0.97 $\pm$ 0.29 & 0.16 $\pm$ 0.15 & 1.40 $\pm$ 0.41 & 1.60 $\pm$ 0.10 & 0.43 & 0 & 20 \\ 
 401857           & e1857 & 16.10 & 0.15 & 4.28 $\pm$ 0.04 & 0.03 $\pm$ 0.01 & 1.40 $\pm$ 0.00 & 2.28 $\pm$ 0.10 & 0.43 & 6 & 6 \\ 
 401857           & e1857 & 16.10 & 0.15 & 3.90 $\pm$ 1.87 & 0.04 $\pm$ 0.07 & 1.40 $\pm$ 0.47 & 1.60 $\pm$ 0.10 & 0.78 & 0 & 11 \\ 
 401925           & e1925 & 18.40 & 0.15 & 0.48 $\pm$ 0.10 & 0.34 $\pm$ 0.19 & 1.40 $\pm$ 0.37 & 1.60 $\pm$ 0.10 & 0.43 & 0 & 5 \\ 
 413123           & f3123 & 19.00 & 0.15 & 1.22 $\pm$ 0.50 & 0.03 $\pm$ 0.05 & 1.40 $\pm$ 0.45 & 1.60 $\pm$ 0.10 & 0.89 & 0 & 29 \\ 
 413123           & f3123 & 19.00 & 0.15 & 1.26 $\pm$ 0.52 & 0.03 $\pm$ 0.08 & 1.40 $\pm$ 0.46 & 1.60 $\pm$ 0.10 & 0.44 & 0 & 12 \\ 
 413192           & f3192 & 16.80 & 0.15 & 2.78 $\pm$ 0.02 & 0.04 $\pm$ 0.01 & 1.40 $\pm$ 0.00 & 1.31 $\pm$ 0.10 & 0.33 & 23 & 25 \\ 
 413192           & f3192 & 16.80 & 0.15 & 2.16 $\pm$ 0.02 & 0.07 $\pm$ 0.01 & 1.40 $\pm$ 0.00 & 1.49 $\pm$ 0.10 & 0.84 & 114 & 118 \\ 
 414287           & f4287 & 17.70 & 0.15 & 1.97 $\pm$ 0.74 & 0.04 $\pm$ 0.04 & 1.40 $\pm$ 0.41 & 1.60 $\pm$ 0.10 & 0.57 & 0 & 9 \\ 
 414772           & f4772 & 19.00 & 0.15 & 1.00 $\pm$ 0.68 & 0.04 $\pm$ 0.06 & 1.00 $\pm$ 0.73 & 1.60 $\pm$ 0.10 & 0.95 & 0 & 5 \\ 
 415711           & f5711 & 19.00 & 0.15 & 0.35 $\pm$ 0.10 & 0.37 $\pm$ 0.26 & 1.40 $\pm$ 0.46 & 1.60 $\pm$ 0.10 & 0.66 & 0 & 9 \\ 
 415986           & f5986 & 18.10 & 0.15 & 1.07 $\pm$ 0.54 & 0.09 $\pm$ 0.11 & 1.40 $\pm$ 0.57 & 1.60 $\pm$ 0.10 & 0.50 & 0 & 10 \\ 
 415986           & f5986 & 18.10 & 0.15 & 1.08 $\pm$ 0.29 & 0.09 $\pm$ 0.08 & 1.40 $\pm$ 0.33 & 1.60 $\pm$ 0.10 & 0.87 & 0 & 27 \\ 
 416071           & f6071 & 17.90 & 0.15 & 0.80 $\pm$ 0.01 & 0.19 $\pm$ 0.03 & 1.40 $\pm$ 0.00 & 2.24 $\pm$ 0.10 & 0.33 & 8 & 8 \\ 
 417264           & f7264 & 17.20 & 0.15 & 1.93 $\pm$ 0.02 & 0.06 $\pm$ 0.01 & 1.40 $\pm$ 0.00 & 2.16 $\pm$ 0.10 & 0.68 & 15 & 15 \\ 
 417264           & f7264 & 17.20 & 0.15 & 1.93 $\pm$ 0.01 & 0.06 $\pm$ 0.01 & 1.40 $\pm$ 0.00 & 2.59 $\pm$ 0.10 & 0.88 & 27 & 27 \\ 
 417264           & f7264 & 17.20 & 0.15 & 2.72 $\pm$ 1.26 & 0.03 $\pm$ 0.04 & 1.40 $\pm$ 0.43 & 1.60 $\pm$ 0.10 & 0.84 & 0 & 11 \\ 
 418797           & f8797 & 19.50 & 0.15 & 0.79 $\pm$ 0.23 & 0.05 $\pm$ 0.06 & 1.40 $\pm$ 0.36 & 1.60 $\pm$ 0.10 & 0.39 & 0 & 7 \\ 
 422699           & g2699 & 18.30 & 0.15 & 0.62 $\pm$ 0.24 & 0.22 $\pm$ 0.18 & 1.40 $\pm$ 0.52 & 1.60 $\pm$ 0.10 & 0.97 & 0 & 11 \\ 
 424089           & g4089 & 17.70 & 0.15 & 2.31 $\pm$ 0.72 & 0.03 $\pm$ 0.03 & 1.40 $\pm$ 0.32 & 1.60 $\pm$ 0.10 & 0.49 & 0 & 11 \\ 
 424392           & g4392 & 21.90 & 0.15 & 0.24 $\pm$ 0.10 & 0.05 $\pm$ 0.06 & 1.40 $\pm$ 0.49 & 1.60 $\pm$ 0.10 & 0.46 & 0 & 9 \\ 
 428223           & g8223 & 16.10 & 0.15 & 2.53 $\pm$ 0.82 & 0.10 $\pm$ 0.13 & 1.40 $\pm$ 0.37 & 1.60 $\pm$ 0.10 & 0.67 & 0 & 10 \\ 
 429746           & g9746 & 17.30 & 0.15 & 1.27 $\pm$ 0.50 & 0.13 $\pm$ 0.10 & 1.40 $\pm$ 0.42 & 1.60 $\pm$ 0.10 & 0.55 & 0 & 8 \\ 
 431107           & h1107 & 17.70 & 0.15 & 1.27 $\pm$ 0.44 & 0.09 $\pm$ 0.15 & 1.40 $\pm$ 0.43 & 1.60 $\pm$ 0.10 & 0.66 & 0 & 8 \\ 
 433953           & h3953 & 20.90 & 0.15 & 0.26 $\pm$ 0.09 & 0.11 $\pm$ 0.13 & 1.40 $\pm$ 0.44 & 1.60 $\pm$ 0.10 & 0.80 & 0 & 7 \\ 
 433992           & h3992 & 18.00 & 0.15 & 0.88 $\pm$ 0.02 & 0.14 $\pm$ 0.02 & 1.40 $\pm$ 0.00 & 2.52 $\pm$ 0.10 & 0.38 & 5 & 5 \\ 
 434096           & h4096 & 18.00 & 0.15 & 0.53 $\pm$ 0.02 & 0.43 $\pm$ 0.08 & 1.40 $\pm$ 0.00 & 2.32 $\pm$ 0.10 & 0.82 & 4 & 6 \\ 
 434633           & h4633 & 20.90 & 0.15 & 0.31 $\pm$ 0.12 & 0.08 $\pm$ 0.09 & 1.40 $\pm$ 0.45 & 1.60 $\pm$ 0.10 & 0.31 & 0 & 6 \\ 
 434633           & h4633 & 20.90 & 0.15 & 0.44 $\pm$ 0.14 & 0.04 $\pm$ 0.02 & 1.40 $\pm$ 0.37 & 1.60 $\pm$ 0.10 & 0.37 & 0 & 13 \\ 
 436671           & h6671 & 18.00 & 0.15 & 2.16 $\pm$ 0.02 & 0.02 $\pm$ 0.01 & 1.40 $\pm$ 0.00 & 4.24 $\pm$ 0.10 & 0.44 & 7 & 7 \\ 
 437879           & h7879 & 17.70 & 0.15 & 2.24 $\pm$ 0.95 & 0.03 $\pm$ 0.06 & 1.40 $\pm$ 0.45 & 1.60 $\pm$ 0.10 & 0.87 & 0 & 13 \\ 
 437994           & h7994 & 17.30 & 0.15 & 0.80 $\pm$ 0.21 & 0.33 $\pm$ 0.16 & 1.40 $\pm$ 0.38 & 1.60 $\pm$ 0.10 & 0.87 & 0 & 11 \\ 
 438990           & h8990 & 18.30 & 0.15 & 0.82 $\pm$ 0.37 & 0.13 $\pm$ 0.16 & 1.40 $\pm$ 0.53 & 1.60 $\pm$ 0.10 & 0.48 & 0 & 5 \\ 
 439889           & h9889 & 20.10 & 0.15 & 0.59 $\pm$ 0.18 & 0.05 $\pm$ 0.04 & 1.40 $\pm$ 0.38 & 1.60 $\pm$ 0.10 & 0.31 & 0 & 4 \\ 
 442605           & i2605 & 19.10 & 0.15 & 0.44 $\pm$ 0.11 & 0.21 $\pm$ 0.17 & 1.40 $\pm$ 0.37 & 1.60 $\pm$ 0.10 & 0.50 & 0 & 5 \\ 
 442742           & i2742 & 17.60 & 0.15 & 2.00 $\pm$ 0.01 & 0.04 $\pm$ 0.01 & 1.40 $\pm$ 0.00 & 2.83 $\pm$ 0.10 & 0.92 & 71 & 78 \\ 
 443806           & i3806 & 22.00 & 0.15 & 0.29 $\pm$ 0.12 & 0.03 $\pm$ 0.04 & 1.40 $\pm$ 0.52 & 1.60 $\pm$ 0.10 & 0.24 & 0 & 12 \\ 
 443880           & i3880 & 19.40 & 0.15 & 0.25 $\pm$ 0.04 & 0.50 $\pm$ 0.17 & 1.80 $\pm$ 0.55 & 1.60 $\pm$ 0.10 & 0.65 & 0 & 6 \\ 
 443923           & i3923 & 17.40 & 0.15 & 2.15 $\pm$ 1.01 & 0.04 $\pm$ 0.07 & 1.00 $\pm$ 0.58 & 1.60 $\pm$ 0.10 & 0.52 & 0 & 11 \\ 
 445025           & i5025 & 17.50 & 0.15 & 2.10 $\pm$ 0.02 & 0.04 $\pm$ 0.01 & 1.40 $\pm$ 0.00 & 0.70 $\pm$ 0.10 & 0.09 & 5 & 5 \\ 
 445305           & i5305 & 19.90 & 0.15 & 0.80 $\pm$ 0.31 & 0.03 $\pm$ 0.05 & 1.40 $\pm$ 0.45 & 1.60 $\pm$ 0.10 & 0.20 & 0 & 4 \\ 
 450159           & j0159 & 18.90 & 0.15 & 0.73 $\pm$ 0.23 & 0.09 $\pm$ 0.10 & 1.40 $\pm$ 0.40 & 1.60 $\pm$ 0.10 & 0.62 & 0 & 5 \\ 
 453687           & j3687 & 19.30 & 0.15 & 1.06 $\pm$ 0.48 & 0.03 $\pm$ 0.03 & 1.40 $\pm$ 0.47 & 1.60 $\pm$ 0.10 & 0.25 & 0 & 13 \\ 
 453707           & j3707 & 18.60 & 0.15 & 0.50 $\pm$ 0.14 & 0.26 $\pm$ 0.12 & 1.40 $\pm$ 0.39 & 1.60 $\pm$ 0.10 & 0.40 & 0 & 4 \\ 
 454078           & j4078 & 17.50 & 0.15 & 2.96 $\pm$ 1.20 & 0.02 $\pm$ 0.04 & 1.40 $\pm$ 0.44 & 1.60 $\pm$ 0.10 & 0.23 & 0 & 15 \\ 
 454100           & j4100 & 20.10 & 0.15 & 0.55 $\pm$ 0.01 & 0.05 $\pm$ 0.01 & 1.40 $\pm$ 0.00 & 1.70 $\pm$ 0.10 & 0.26 & 6 & 6 \\ 
 2002 GP186           & K02GI6P & 20.30 & 0.15 & 0.17 $\pm$ 0.05 & 0.44 $\pm$ 0.27 & 1.40 $\pm$ 0.48 & 1.60 $\pm$ 0.10 & 0.29 & 0 & 5 \\ 
 2003  KZ18           & K03K18Z & 21.20 & 0.15 & 0.47 $\pm$ 0.19 & 0.03 $\pm$ 0.04 & 1.40 $\pm$ 0.48 & 1.60 $\pm$ 0.10 & 0.69 & 0 & 6 \\ 
 2003   MT9           & K03M09T & 18.60 & 0.15 & 0.68 $\pm$ 0.22 & 0.14 $\pm$ 0.17 & 1.40 $\pm$ 0.46 & 1.60 $\pm$ 0.10 & 0.30 & 0 & 4 \\ 
 2006  KL89           & K06K89L & 18.60 & 0.15 & 0.96 $\pm$ 0.38 & 0.07 $\pm$ 0.03 & 1.40 $\pm$ 0.41 & 1.60 $\pm$ 0.10 & 0.68 & 0 & 23 \\ 
 2006  KL89           & K06K89L & 18.60 & 0.15 & 1.07 $\pm$ 0.52 & 0.06 $\pm$ 0.10 & 1.40 $\pm$ 0.53 & 1.60 $\pm$ 0.10 & 0.65 & 0 & 15 \\ 
 2006   OF5           & K06O05F & 19.30 & 0.15 & 0.95 $\pm$ 0.55 & 0.04 $\pm$ 0.03 & 1.00 $\pm$ 0.64 & 1.60 $\pm$ 0.10 & 0.63 & 0 & 46 \\ 
 2006   OF5           & K06O05F & 19.30 & 0.15 & 0.93 $\pm$ 0.32 & 0.04 $\pm$ 0.03 & 1.00 $\pm$ 0.49 & 1.60 $\pm$ 0.10 & 0.43 & 0 & 47 \\ 
 2006 UR217           & K06UL7R & 19.80 & 0.15 & 0.89 $\pm$ 0.07 & 0.03 $\pm$ 0.01 & 1.40 $\pm$ 0.11 & 6.16 $\pm$ 0.10 & 0.26 & 8 & 9 \\ 
 2006 UR217           & K06UL7R & 19.80 & 0.15 & 0.95 $\pm$ 0.40 & 0.02 $\pm$ 0.05 & 1.40 $\pm$ 0.47 & 1.60 $\pm$ 0.10 & 0.44 & 0 & 6 \\ 
 2007  WE55           & K07W55E & 20.20 & 0.15 & 0.69 $\pm$ 0.18 & 0.03 $\pm$ 0.05 & 1.40 $\pm$ 0.33 & 1.60 $\pm$ 0.10 & 0.78 & 0 & 13 \\ 
 2010   CO1           & K10C01O & 21.80 & 0.15 & 0.29 $\pm$ 0.15 & 0.04 $\pm$ 0.06 & 1.00 $\pm$ 0.67 & 1.60 $\pm$ 0.10 & 0.43 & 0 & 13 \\ 
 2010  LF86           & K10L86F & 17.20 & 0.15 & 2.56 $\pm$ 1.22 & 0.04 $\pm$ 0.05 & 1.40 $\pm$ 0.50 & 1.60 $\pm$ 0.10 & 0.49 & 0 & 5 \\ 
 2010   UB8           & K10U08B & 19.60 & 0.15 & 0.92 $\pm$ 0.08 & 0.03 $\pm$ 0.01 & 1.16 $\pm$ 0.09 & 1.60 $\pm$ 0.10 & 0.16 & 7 & 7 \\ 
 2010   UB8           & K10U08B & 19.60 & 0.15 & 0.88 $\pm$ 0.42 & 0.03 $\pm$ 0.05 & 1.00 $\pm$ 0.61 & 1.60 $\pm$ 0.10 & 0.37 & 0 & 15 \\ 
 2010   YD3           & K10Y03D & 20.00 & 0.15 & 0.76 $\pm$ 0.35 & 0.03 $\pm$ 0.04 & 1.40 $\pm$ 0.52 & 1.60 $\pm$ 0.10 & 0.54 & 0 & 12 \\ 
 2011  AM24           & K11A24M & 20.50 & 0.15 & 0.50 $\pm$ 0.01 & 0.04 $\pm$ 0.01 & 1.40 $\pm$ 0.01 & 0.47 $\pm$ 0.10 & 0.49 & 8 & 8 \\ 
 2011  AM24           & K11A24M & 20.50 & 0.15 & 0.51 $\pm$ 0.01 & 0.04 $\pm$ 0.01 & 1.40 $\pm$ 0.01 & 0.94 $\pm$ 0.10 & 0.47 & 14 & 15 \\ 
 2011  HJ61           & K11H61J & 19.30 & 0.15 & 1.28 $\pm$ 0.57 & 0.02 $\pm$ 0.04 & 1.40 $\pm$ 0.47 & 1.60 $\pm$ 0.10 & 0.97 & 0 & 13 \\ 
 2011   JU2           & K11J02U & 18.40 & 0.15 & 1.49 $\pm$ 0.56 & 0.03 $\pm$ 0.06 & 1.40 $\pm$ 0.42 & 1.60 $\pm$ 0.10 & 0.77 & 0 & 4 \\ 
 2011   OL5           & K11O05L & 20.20 & 0.15 & 0.28 $\pm$ 0.08 & 0.19 $\pm$ 0.16 & 1.40 $\pm$ 0.42 & 1.60 $\pm$ 0.10 & 1.05 & 0 & 36 \\ 
 2011   OL5           & K11O05L & 20.20 & 0.15 & 0.28 $\pm$ 0.11 & 0.19 $\pm$ 0.16 & 1.40 $\pm$ 0.51 & 1.60 $\pm$ 0.10 & 0.60 & 0 & 13 \\ 
 2011   VQ5           & K11V05Q & 20.10 & 0.15 & 0.56 $\pm$ 0.23 & 0.05 $\pm$ 0.08 & 1.40 $\pm$ 0.50 & 1.60 $\pm$ 0.10 & 0.85 & 0 & 18 \\ 
 2011  YB40           & K11Y40B & 19.10 & 0.15 & 0.42 $\pm$ 0.12 & 0.22 $\pm$ 0.17 & 1.40 $\pm$ 0.42 & 1.60 $\pm$ 0.10 & 0.38 & 0 & 5 \\ 
 2012   OD1           & K12O01D & 18.60 & 0.15 & 0.35 $\pm$ 0.09 & 0.54 $\pm$ 0.26 & 1.40 $\pm$ 0.44 & 1.60 $\pm$ 0.10 & 0.36 & 0 & 5 \\ 
 2014  JY24           & K14J24Y & 18.30 & 0.15 & 1.92 $\pm$ 1.01 & 0.02 $\pm$ 0.05 & 1.40 $\pm$ 0.55 & 1.60 $\pm$ 0.10 & 0.37 & 0 & 15 \\ 
 2014 QK434           & K14Qh4K & 19.10 & 0.15 & 0.30 $\pm$ 0.01 & 0.46 $\pm$ 0.05 & 1.40 $\pm$ 0.00 & 0.83 $\pm$ 0.10 & 0.41 & 5 & 5 \\ 
 2014  TA36           & K14T36A & 20.70 & 0.15 & 0.55 $\pm$ 0.23 & 0.03 $\pm$ 0.03 & 1.40 $\pm$ 0.49 & 1.60 $\pm$ 0.10 & 0.69 & 0 & 11 \\ 
 2014    US           & K14U00S & 19.10 & 0.15 & 0.47 $\pm$ 0.15 & 0.19 $\pm$ 0.26 & 1.40 $\pm$ 0.46 & 1.60 $\pm$ 0.10 & 0.30 & 0 & 16 \\ 
 2014    US           & K14U00S & 19.10 & 0.15 & 0.56 $\pm$ 0.19 & 0.13 $\pm$ 0.08 & 1.40 $\pm$ 0.42 & 1.60 $\pm$ 0.10 & 0.38 & 0 & 14 \\ 
 2014  UV33           & K14U33V & 17.90 & 0.15 & 0.82 $\pm$ 0.02 & 0.18 $\pm$ 0.03 & 1.40 $\pm$ 0.00 & 3.29 $\pm$ 0.10 & 1.58 & 20 & 24 \\ 
 2014 UF206           & K14UK6F & 18.80 & 0.15 & 1.52 $\pm$ 0.68 & 0.02 $\pm$ 0.02 & 1.00 $\pm$ 0.53 & 1.60 $\pm$ 0.10 & 0.68 & 0 & 46 \\ 
 2014 UF206           & K14UK6F & 18.80 & 0.15 & 1.29 $\pm$ 0.06 & 0.03 $\pm$ 0.01 & 1.12 $\pm$ 0.05 & 1.60 $\pm$ 0.10 & 0.16 & 23 & 23 \\ 
 2014 WF365           & K14Wa5F & 17.20 & 0.15 & 2.18 $\pm$ 0.02 & 0.05 $\pm$ 0.01 & 1.40 $\pm$ 0.00 & 1.39 $\pm$ 0.10 & 0.20 & 7 & 7 \\ 
 2014   XR6           & K14X06R & 18.30 & 0.15 & 0.86 $\pm$ 0.29 & 0.11 $\pm$ 0.17 & 1.40 $\pm$ 0.45 & 1.60 $\pm$ 0.10 & 0.56 & 0 & 10 \\ 
 2014  XX31           & K14X31X & 17.50 & 0.15 & 1.49 $\pm$ 0.53 & 0.08 $\pm$ 0.06 & 1.40 $\pm$ 0.39 & 1.60 $\pm$ 0.10 & 0.67 & 0 & 6 \\ 
 2014  YJ14           & K14Y14J & 18.30 & 0.15 & 1.91 $\pm$ 0.12 & 0.02 $\pm$ 0.01 & 1.12 $\pm$ 0.06 & 1.60 $\pm$ 0.10 & 0.12 & 7 & 7 \\ 
 2014  YS14           & K14Y14S & 21.10 & 0.15 & 0.30 $\pm$ 0.12 & 0.07 $\pm$ 0.09 & 1.40 $\pm$ 0.51 & 1.60 $\pm$ 0.10 & 0.57 & 0 & 5 \\ 
 2014  YT14           & K14Y14T & 18.90 & 0.15 & 1.16 $\pm$ 0.47 & 0.04 $\pm$ 0.02 & 1.00 $\pm$ 0.52 & 1.60 $\pm$ 0.10 & 0.95 & 0 & 18 \\ 
 2014  YS34           & K14Y34S & 20.80 & 0.15 & 0.13 $\pm$ 0.03 & 0.50 $\pm$ 0.23 & 1.40 $\pm$ 0.41 & 1.60 $\pm$ 0.10 & 0.46 & 0 & 5 \\ 
 2014  YB35           & K14Y35B & 19.00 & 0.15 & 0.28 $\pm$ 0.01 & 0.57 $\pm$ 0.07 & 1.40 $\pm$ 0.00 & 1.31 $\pm$ 0.10 & 0.52 & 6 & 6 \\ 
 2014  YR43           & K14Y43R & 19.50 & 0.15 & 0.37 $\pm$ 0.13 & 0.20 $\pm$ 0.13 & 1.40 $\pm$ 0.47 & 1.60 $\pm$ 0.10 & 0.36 & 0 & 9 \\ 
 2015  AC17           & K15A17C & 19.90 & 0.15 & 0.67 $\pm$ 0.28 & 0.04 $\pm$ 0.04 & 1.00 $\pm$ 0.57 & 1.60 $\pm$ 0.10 & 0.67 & 0 & 34 \\ 
 2015 AY245           & K15AO5Y & 21.20 & 0.15 & 0.37 $\pm$ 0.03 & 0.04 $\pm$ 0.02 & 1.40 $\pm$ 0.12 & 8.64 $\pm$ 0.10 & 0.43 & 13 & 13 \\ 
 2015 AY245           & K15AO5Y & 21.20 & 0.15 & 0.39 $\pm$ 0.18 & 0.04 $\pm$ 0.09 & 1.40 $\pm$ 0.59 & 1.60 $\pm$ 0.10 & 2.84 & 0 & 60 \\ 
 2015 AK280           & K15AS0K & 21.80 & 0.15 & 0.36 $\pm$ 0.12 & 0.03 $\pm$ 0.04 & 1.40 $\pm$ 0.41 & 1.60 $\pm$ 0.10 & 0.36 & 0 & 4 \\ 
 2015 BY516           & K15Bp6Y & 22.30 & 0.15 & 0.24 $\pm$ 0.12 & 0.04 $\pm$ 0.03 & 1.00 $\pm$ 0.64 & 1.60 $\pm$ 0.10 & 0.55 & 0 & 8 \\ 
 2015  CV13           & K15C13V & 20.30 & 0.15 & 0.44 $\pm$ 0.13 & 0.07 $\pm$ 0.04 & 1.40 $\pm$ 0.35 & 1.60 $\pm$ 0.10 & 0.69 & 0 & 8 \\ 
 2015 DE176           & K15DH6E & 19.70 & 0.15 & 0.68 $\pm$ 0.04 & 0.05 $\pm$ 0.01 & 1.40 $\pm$ 0.07 & 0.21 $\pm$ 0.10 & 0.19 & 9 & 9 \\ 
 2015 DE176           & K15DH6E & 19.70 & 0.15 & 0.57 $\pm$ 0.29 & 0.07 $\pm$ 0.11 & 1.40 $\pm$ 0.59 & 1.60 $\pm$ 0.10 & 0.25 & 0 & 4 \\ 
 2015 DX198           & K15DJ8X & 22.00 & 0.15 & 0.35 $\pm$ 0.10 & 0.02 $\pm$ 0.02 & 1.40 $\pm$ 0.38 & 1.60 $\pm$ 0.10 & 1.31 & 0 & 9 \\ 
 2015    EZ           & K15E00Z & 20.30 & 0.15 & 0.19 $\pm$ 0.05 & 0.36 $\pm$ 0.19 & 1.40 $\pm$ 0.39 & 1.60 $\pm$ 0.10 & 0.62 & 0 & 10 \\ 
 2015  FZ35           & K15F35Z & 19.40 & 0.15 & 0.64 $\pm$ 0.21 & 0.08 $\pm$ 0.11 & 1.40 $\pm$ 0.41 & 1.60 $\pm$ 0.10 & 0.57 & 0 & 5 \\ 
 2015 FY117           & K15FB7Y & 21.30 & 0.15 & 0.38 $\pm$ 0.17 & 0.04 $\pm$ 0.04 & 1.00 $\pm$ 0.60 & 1.60 $\pm$ 0.10 & 0.62 & 0 & 43 \\ 
 2015 FH120           & K15FC0H & 18.70 & 0.15 & 0.75 $\pm$ 0.26 & 0.11 $\pm$ 0.11 & 1.40 $\pm$ 0.46 & 1.60 $\pm$ 0.10 & 0.25 & 0 & 11 \\ 
 2015 FU332           & K15FX2U & 17.20 & 0.15 & 0.94 $\pm$ 0.36 & 0.26 $\pm$ 0.27 & 1.40 $\pm$ 0.53 & 1.60 $\pm$ 0.10 & 0.47 & 0 & 9 \\ 
 2015 FD341           & K15FY1D & 17.70 & 0.15 & 1.25 $\pm$ 0.48 & 0.09 $\pm$ 0.08 & 1.40 $\pm$ 0.43 & 1.60 $\pm$ 0.10 & 0.77 & 0 & 4 \\ 
 2015 FT344           & K15FY4T & 19.90 & 0.15 & 0.75 $\pm$ 0.24 & 0.03 $\pm$ 0.03 & 1.40 $\pm$ 0.39 & 1.60 $\pm$ 0.10 & 0.60 & 0 & 4 \\ 
 2015 FT344           & K15FY4T & 19.90 & 0.15 & 0.76 $\pm$ 0.21 & 0.03 $\pm$ 0.04 & 1.40 $\pm$ 0.34 & 1.60 $\pm$ 0.10 & 0.46 & 0 & 7 \\ 
 2015    GY           & K15G00Y & 21.70 & 0.15 & 0.14 $\pm$ 0.05 & 0.18 $\pm$ 0.19 & 1.40 $\pm$ 0.50 & 1.60 $\pm$ 0.10 & 0.24 & 0 & 4 \\ 
 2015  GK50           & K15G50K & 20.60 & 0.15 & 0.61 $\pm$ 0.30 & 0.03 $\pm$ 0.03 & 1.00 $\pm$ 0.62 & 1.60 $\pm$ 0.10 & 0.48 & 0 & 9 \\ 
 2015  GK50           & K15G50K & 20.60 & 0.15 & 0.46 $\pm$ 0.03 & 0.05 $\pm$ 0.01 & 0.99 $\pm$ 0.06 & 1.60 $\pm$ 0.10 & 0.27 & 16 & 16 \\ 
 2015  GN50           & K15G50N & 20.20 & 0.15 & 0.29 $\pm$ 0.11 & 0.18 $\pm$ 0.12 & 1.40 $\pm$ 0.46 & 1.60 $\pm$ 0.10 & 0.36 & 0 & 5 \\ 
 2015  HF11           & K15H11F & 19.40 & 0.15 & 1.11 $\pm$ 0.44 & 0.02 $\pm$ 0.01 & 1.40 $\pm$ 0.42 & 1.60 $\pm$ 0.10 & 0.82 & 0 & 10 \\ 
 2015  JF11           & K15J11F & 21.20 & 0.15 & 0.17 $\pm$ 0.05 & 0.20 $\pm$ 0.11 & 1.40 $\pm$ 0.40 & 1.60 $\pm$ 0.10 & 0.34 & 0 & 6 \\ 
 2015 KH157           & K15KF7H & 20.00 & 0.15 & 0.58 $\pm$ 0.23 & 0.05 $\pm$ 0.11 & 1.40 $\pm$ 0.49 & 1.60 $\pm$ 0.10 & 0.22 & 0 & 9 \\ 
 2015 KL157           & K15KF7L & 19.10 & 0.15 & 1.45 $\pm$ 0.31 & 0.02 $\pm$ 0.01 & 1.00 $\pm$ 0.38 & 1.50 $\pm$ 0.10 & 0.76 & 0 & 48 \\ 
 2015 KL157           & K15KF7L & 19.10 & 0.15 & 0.36 $\pm$ 0.01 & 0.30 $\pm$ 0.05 & 0.40 $\pm$ 0.00 & 1.50 $\pm$ 0.10 & 1.05 & 9 & 12 \\ 
 2015  LK24           & K15L24K & 21.60 & 0.15 & 0.31 $\pm$ 0.11 & 0.04 $\pm$ 0.07 & 1.40 $\pm$ 0.44 & 1.60 $\pm$ 0.10 & 0.39 & 0 & 8 \\ 
 2015 MQ130           & K15MD0Q & 20.90 & 0.15 & 0.36 $\pm$ 0.07 & 0.06 $\pm$ 0.03 & 0.54 $\pm$ 0.10 & 1.60 $\pm$ 0.10 & 0.36 & 5 & 7 \\ 
 2015 MQ130           & K15MD0Q & 20.90 & 0.15 & 0.55 $\pm$ 0.26 & 0.03 $\pm$ 0.03 & 1.00 $\pm$ 0.61 & 1.60 $\pm$ 0.10 & 0.83 & 0 & 4 \\ 
 2015  NA14           & K15N14A & 22.00 & 0.15 & 0.09 $\pm$ 0.02 & 0.34 $\pm$ 0.18 & 1.40 $\pm$ 0.32 & 1.60 $\pm$ 0.10 & 0.45 & 0 & 5 \\ 
 2015  OA22           & K15O22A & 20.00 & 0.15 & 0.79 $\pm$ 0.34 & 0.03 $\pm$ 0.03 & 1.00 $\pm$ 0.56 & 1.60 $\pm$ 0.10 & 0.11 & 0 & 4 \\ 
 2015  OS35           & K15O35S & 19.10 & 0.15 & 1.26 $\pm$ 0.01 & 0.03 $\pm$ 0.00 & 1.40 $\pm$ 0.00 & 5.95 $\pm$ 0.10 & 0.17 & 20 & 22 \\ 
 2015  OS35           & K15O35S & 19.10 & 0.15 & 1.42 $\pm$ 0.01 & 0.02 $\pm$ 0.00 & 1.40 $\pm$ 0.00 & 6.34 $\pm$ 0.10 & 0.15 & 10 & 10 \\ 
 2015    PD           & K15P00D & 19.30 & 0.15 & 0.62 $\pm$ 0.22 & 0.09 $\pm$ 0.10 & 1.40 $\pm$ 0.44 & 1.60 $\pm$ 0.10 & 0.57 & 0 & 8 \\ 
 2015  PM57           & K15P57M & 18.60 & 0.15 & 0.59 $\pm$ 0.21 & 0.19 $\pm$ 0.22 & 1.40 $\pm$ 0.49 & 1.60 $\pm$ 0.10 & 0.22 & 0 & 7 \\ 
 2015   QM3           & K15Q03M & 20.30 & 0.15 & 0.27 $\pm$ 0.05 & 0.19 $\pm$ 0.15 & 1.40 $\pm$ 0.27 & 1.60 $\pm$ 0.10 & 0.61 & 0 & 9 \\ 
 2015  RS83           & K15R83S & 19.40 & 0.15 & 0.47 $\pm$ 0.13 & 0.14 $\pm$ 0.13 & 1.40 $\pm$ 0.37 & 1.60 $\pm$ 0.10 & 0.52 & 0 & 5 \\ 
 2015 RR150           & K15RF0R & 19.70 & 0.15 & 0.34 $\pm$ 0.12 & 0.20 $\pm$ 0.19 & 1.40 $\pm$ 0.46 & 1.60 $\pm$ 0.10 & 0.60 & 0 & 13 \\ 
 2015  SF20           & K15S20F & 19.70 & 0.15 & 0.40 $\pm$ 0.16 & 0.15 $\pm$ 0.14 & 1.40 $\pm$ 0.53 & 1.60 $\pm$ 0.10 & 0.52 & 0 & 8 \\ 
 2015  SS20           & K15S20S & 22.40 & 0.15 & 0.26 $\pm$ 0.10 & 0.03 $\pm$ 0.03 & 1.40 $\pm$ 0.46 & 1.60 $\pm$ 0.10 & 0.44 & 0 & 7 \\ 
 2015 TK237           & K15TN7K & 22.60 & 0.15 & 0.23 $\pm$ 0.07 & 0.03 $\pm$ 0.04 & 1.40 $\pm$ 0.40 & 1.60 $\pm$ 0.10 & 0.53 & 0 & 7 \\ 
 2015 TW346           & K15TY6W & 18.60 & 0.15 & 1.26 $\pm$ 0.46 & 0.04 $\pm$ 0.04 & 1.40 $\pm$ 0.41 & 1.60 $\pm$ 0.10 & 0.44 & 0 & 15 \\ 
 2015  UK52           & K15U52K & 20.10 & 0.15 & 0.21 $\pm$ 0.06 & 0.35 $\pm$ 0.22 & 1.40 $\pm$ 0.45 & 1.60 $\pm$ 0.10 & 0.34 & 0 & 5 \\ 
 2015 VZ145           & K15VE5Z & 23.70 & 0.15 & 0.16 $\pm$ 0.06 & 0.02 $\pm$ 0.05 & 1.40 $\pm$ 0.52 & 1.60 $\pm$ 0.10 & 1.36 & 0 & 5 \\ 
 2015  WM16           & K15W16M & 21.80 & 0.15 & 0.39 $\pm$ 0.06 & 0.02 $\pm$ 0.01 & 1.10 $\pm$ 0.16 & 1.60 $\pm$ 0.10 & 0.17 & 8 & 8 \\ 
 2015 XB130           & K15XD0B & 21.80 & 0.15 & 0.33 $\pm$ 0.13 & 0.03 $\pm$ 0.04 & 1.40 $\pm$ 0.48 & 1.60 $\pm$ 0.10 & 0.58 & 0 & 9 \\ 
 2015 XY378           & K15Xb8Y & 19.60 & 0.15 & 0.31 $\pm$ 0.11 & 0.26 $\pm$ 0.23 & 1.40 $\pm$ 0.46 & 1.60 $\pm$ 0.10 & 0.32 & 0 & 5 \\ 
 
\enddata
\label{tab:neoyr2}
\end{deluxetable}
 
\begin{deluxetable}{rrrrrrrrrrr}
\tabletypesize{\scriptsize}
\tablecaption{Measured diameters ($d$) and albedos ($p_V$) of non-NEA asteroids observed during Year 2 of the NEOWISE Reactivation mission. Asteroids may be identified by numbers, provisional designations, or via the MPC packed format. Beaming $\eta$, $H$, $G$, the amplitude of the 4.6 $\mu$m light curve (W2 amp., in mag), and the numbers of observations used in the 3.4 $\mu$m ($n_{W1}$) and 4.6 $\mu$m ($n_{W2}$) wavelengths are also reported. 
Only the first 10 lines are shown; the remainder are available in electronic format through the journal website. }
\tablewidth{0pt}
\tablehead{
	\colhead{Object} & \colhead{Packed} & \colhead{$H$}& \colhead{$G$}& \colhead{$d$ (km)} & \colhead{$p_V$} & \colhead{$\eta$}& \colhead{$p_{IR}/p_{V}$} & \colhead{W2 amp.} & \colhead{$n_{W1}$} & \colhead{$n_{W2}$} }                                                                                        
\startdata
 10           & 00010 & 5.46 & 0.12 & 450.53 $\pm$ 200.23 & 0.05 $\pm$ 0.05 & 0.95 $\pm$ 0.23 & 1.50 $\pm$ 0.10 & 0.05 & 4 & 4 \\ 
 13           & 00013 & 6.77 & 0.12 & 207.98 $\pm$ 46.68 & 0.08 $\pm$ 0.04 & 1.00 $\pm$ 0.35 & 1.00 $\pm$ 0.60 & 0.16 & 5 & 5 \\ 
 13           & 00013 & 6.77 & 0.12 & 192.79 $\pm$ 53.36 & 0.09 $\pm$ 0.06 & 1.00 $\pm$ 0.39 & 1.00 $\pm$ 0.60 & 0.32 & 9 & 9 \\ 
 19           & 00019 & 7.20 & 0.12 & 176.97 $\pm$ 56.71 & 0.06 $\pm$ 0.07 & 0.95 $\pm$ 0.21 & 1.50 $\pm$ 0.10 & 0.25 & 10 & 10 \\ 
 19           & 00019 & 7.20 & 0.12 & 182.71 $\pm$ 40.61 & 0.06 $\pm$ 0.03 & 0.95 $\pm$ 0.14 & 1.50 $\pm$ 0.10 & 0.32 & 8 & 8 \\ 
 21           & 00021 & 7.45 & 0.24 & 102.07 $\pm$ 24.56 & 0.18 $\pm$ 0.08 & 1.00 $\pm$ 0.39 & 1.00 $\pm$ 0.60 & 0.50 & 11 & 11 \\ 
 21           & 00021 & 7.45 & 0.24 & 99.71 $\pm$ 22.62 & 0.18 $\pm$ 0.05 & 1.00 $\pm$ 0.38 & 1.00 $\pm$ 0.60 & 0.35 & 12 & 13 \\ 
 23           & 00023 & 7.09 & 0.24 & 93.99 $\pm$ 20.14 & 0.29 $\pm$ 0.14 & 0.95 $\pm$ 0.20 & 1.50 $\pm$ 0.10 & 0.23 & 9 & 9 \\ 
 30           & 00030 & 7.67 & 0.24 & 105.70 $\pm$ 23.25 & 0.19 $\pm$ 0.11 & 0.95 $\pm$ 0.19 & 1.50 $\pm$ 0.10 & 0.23 & 6 & 7 \\ 
 33           & 00033 & 8.60 & 0.24 & 54.39 $\pm$ 11.84 & 0.23 $\pm$ 0.13 & 0.95 $\pm$ 0.19 & 1.50 $\pm$ 0.10 & 0.36 & 7 & 7 \\ 
\enddata
\label{tab:mbanew}
\end{deluxetable}
\begin{deluxetable}{rllll}
\tabletypesize{\scriptsize}
\tablecaption{Measured diameters and albedos for objects observed during the Year 2 Reactivation mission that may be accessible by spacecraft, following the NHATS criteria. Also listed is the minimum mission round trip time in days for each object from \citet{BarbeeNHATS}. }
\tablewidth{0pt}
\tablehead{
\colhead{Number} & \colhead{Designation} & \colhead{$D$ (km)} & \colhead{$p_V$} &  \colhead{Minimum round trip (days)} 
}                                                                                    
\startdata
(35107)  & 1991 VH    & $0.91 \pm 0.03$ & $0.33 \pm 0.04$ & 354 \\ 
(163899) & 2003 SD220 & $0.80 \pm 0.02$ & $0.31 \pm 0.04$ & 122 \\
(363505) & 2003 UC20  & $1.88 \pm 0.01$ & $0.03 \pm 0.00$ & 290 \\ 
(424392) & 2007 YJ    & $0.24 \pm 0.10$ & $0.05 \pm 0.06$ &  98 \\
         & 2011 AM24  & $0.50 \pm 0.01$ & $0.04 \pm 0.01$ & 130 \\
         & 2015 GY    & $0.14 \pm 0.05$ & $0.18 \pm 0.19$ & 346 \\
         & 2015 NA14  & $0.09 \pm 0.02$ & $0.34 \pm 0.18$ & 170 \\
\enddata
\label{tab:NHATS}
\end{deluxetable}

\end{document}